\documentclass[graybox]{svmult}
\usepackage{epsfig}

\usepackage{latexsym,amssymb,amsmath}
\usepackage{mathptmx}       
\usepackage{helvet}         
\usepackage{courier}        
\usepackage{type1cm}        
\usepackage{makeidx}         
\usepackage{graphicx}        
\usepackage{multicol}        
\usepackage[bottom]{footmisc}

\newcommand{\be}{\begin{eqnarray}}
\newcommand{\ee}{\end{eqnarray}}

\newcommand{\ket}[1]{\mbox{$\mid #1\,\rangle$}}

\newcommand{\pro}[2]{\mbox{$\langle\, #1 \mid #2\,\rangle$}}
\newcommand{\expec}[1]{\mbox{$\langle\, #1\,\rangle$}}

\renewcommand{\d}{\mbox{${\rm d}$}}

\newcommand{\rh}{R_{\rm H}}
\newcommand{\mpl}{M_{\rm P}} 
\newcommand{\mg}{M_{\rm G}}
\newcommand{\lp}{\ell_{\rm P}}

\makeindex             

\begin{document}
\title*{Minimum length effects in black hole physics}
\author{Roberto Casadio, Octavian Micu, Piero Nicolini}
\institute{Roberto Casadio
\at Dipartimento di Fisica e Astronomia, Universit\`a di Bologna,
and I.N.F.N., Sezione di Bologna, via~Irnerio~46, 40126~Bologna, Italy
\email{casadio@bo.infn.it}
\and Octavian Micu
\at Institute of Space Science, Bucharest,
P.O.~Box MG-23, RO-077125 Bucharest-Magurele, Romania
\email{octavian.micu@spacescience.ro}
\and
Piero Nicolini
\at Frankfurt Institute for Advanced Studies (FIAS),
Science Campus Riedberg, Ruth-Moufang-Strasse~1,
60438~Frankfurt am Main, Germany
\email{nicolini@fias.uni-frankfurt.de}}
%
%
\maketitle
\abstract*{We review the main consequences of the possible
existence of a minimum measurable length, of the order of the Planck scale,
on quantum effects occurring in black hole physics.
In particular, we focus on the ensuing minimum mass for black holes
and how modified dispersion relations affect the Hawking decay,
both in four space-time dimensions and in models with extra spatial
dimensions.
In the latter case, we briefly discuss possible phenomenological signatures.
}
\abstract{We review the main consequences of the possible
existence of a minimum measurable length, of the order of the Planck scale,
on quantum effects occurring in black hole physics.
In particular, we focus on the ensuing minimum mass for black holes
and how modified dispersion relations affect the Hawking decay,
both in four space-time dimensions and in models with extra spatial
dimensions.
In the latter case, we briefly discuss possible phenomenological signatures.
}
\section{Gravity and minimum length}
\label{sec:intro}
Physics is characterized by a variety of research fields  and diversified tools of investigation that
strongly depend on the length scales under consideration.
As a result, one finds an array of sub-disciplines, spanning from cosmology, to astrophysics,
geophysics, molecular and atomic physics, nuclear and particle physics. 
In a nutshell, we can say that Physics concerns events occurring at scales between the
radius of the Universe and the typical size of observed elementary particles.
It is not hard to understand that such a rich array of physical phenomena requires specific
formalisms.
For instance, at macroscopic scales, models of the Universe are obtained in terms of
general relativity, while at microscopic scales quantum physics has been proven to be the
adequate theory for the miniaturised world.

Despite the generality of such a scheme, it cannot be considered as complete.
One may be tempted to conclude that at microscopic scales, we can, at least in principle,
figure out arbitrarily small lengths.
In quantum mechanics, or more precisely in quantum field theory, particle sizes are described
by the Compton wavelength, which accounts for the Heisenberg uncertainty in localising
a microscopic object at a given energy.
From this, it descends a ``rule of thumb'' according to which the higher the energy,
the smaller is the size one can probe in a particle physics experiment.
Apparently there is no minimal length scale.
The limitations to the accuracy in measuring a length seems to be only a technological
problem related to the possibility of reaching higher and higher energy scales.   

In the reasoning above, however, we give for granted that quantum physics can be considered
unmodified at any energy scale.
On the contrary, we should better say that the standard quantum formalism is valid for studying
particles and fundamental interactions, provided one of these interactions, gravity, does not produce
relevant effects at the energy under consideration.
The weakness of gravity allows quantum mechanics and quantum field theory to be
efficient tools in a vast variety of physical situations.
On the other hand, already for the Heisenberg microscope, namely a thought experiment
concerning a particle illuminated by light, one should take into account the gravitational
interaction between the particle and the effective mass associated with the energy of the photon.
In doing so, one can discover an additional position uncertainty due to the acceleration the particle
is subject to~\cite{Adl10}.
Accordingly, one has to conclude that there exists a \textit{minimal length}  limiting the accuracy
in localising the particle itself~\cite{KMM95}.
Not surprisingly, such a minimal length depends on the gravitational coupling $G$ and
it is defined as the Planck length through the relation $\lp\equiv \sqrt{\hbar\,G/c^3}\sim 10^{-35}\,$m. 

The Planck scale discloses other important features.
Matter compression (\textit{e.g.} by smashing particles in colliders) is limited by the gravitational
collapse to a black hole, whose size turns out to be the Planck length~\cite{ACS01}.
A further increase of energy in the collision would not lead to a smaller particle but rather to a bigger
black hole, its radius being proportional to the mass.
As a result, the Planck length is not only the smallest scale in particle physics, but also the smallest
admissible size of a black hole.
From this viewpoint, black holes can be reinterpreted as a new ``phase'' of
matter~\cite{ACV08,DGG11,AuS13b} occurring at energies exceeding the Planck mass,
\textit{i.e.}, $\mpl\equiv \hbar\,c^{-1}\,\lp^{-1}= \sqrt{\hbar\,c/G}$.
This general idea is what lies behind the generalised uncertainty principle (GUP),
which we shall derive in a rather new fashion in section~\ref{ssec:hoopBH}
(see Ref.~\cite{Hos12} for a fairly comprehensive account of other derivations).

Despite the simplifications, the idea of a minimal length is supported by all major
formulations of quantum gravity, \textit{i.e.}, an ongoing attempt to formulate a consistent
quantum theory of the gravitational interaction.
Specifically, the existence of a minimum length was evident since the early contribution to
quantum gravity~\cite{DeW62}.
It was clear that space-time has to change its nature when probed at energies of the order
of the Planck scale: rather than a smooth differential manifold, the space-time,
in its quantum regime, is expected to be a fluctuating, foamy structure plagued
by loss of local resolution.
This scenario is confirmed by Planckian scattering of strings~\cite{ACV87},
whose theory can be interpreted as a ``field theory'' of extended objects
encoding a quantum of length $\lambda=\sqrt{\hbar\,\alpha^\prime}$~\cite{Ven86,ACV89,Yon89,KPP90,AuS13}.
In the context of loop quantum gravity, a minimum resolution length emerges from the discreteness
of the spectra of area and volume operators~\cite{RoS95}.
The idea of implementing a gravitational ultraviolet cutoff has followed several routes
like string inspired non-commutative geometry~\cite{SeW99} or asymptotically safe gravity~\cite{Wei80}.
According to the latter proposal, the gravitational interaction becomes weaker and weaker
as the energy increases:
in the  ultraviolet regime there is a non-trivial fixed point at which the theory is safe from
divergences~\cite{Reu96}.
Further analyses of the emergence of a minimal length in quantum gravity can be found
in Ref.~\cite{Ame94,Gar95} (for recent reviews see~\cite{Hos12,SNB12}).

In the following sections we present the relationship between black holes
and a minimum resolution length.
Such a relationship is dual:
we have already mentioned that the size of Planck scale black holes provides
a natural ultraviolet cutoff;
on the other hand, it is instructive to explore the complementary possibility,
namely the study of modifications of classical black hole metrics that we expect
by assuming that the space-time is endowed with a quantum gravity induced
minimal length. 
\section{Minimum black hole mass}
\label{sec:BH}
Before we tackle the issue of the existence of a minimum black hole mass,
let us briefly review the key ingredient that suggests it is sensible to put
together Heisenberg's fundamental uncertainty principle and a gravitational
source of error, thus yielding a GUP and a minimum measurable
length~\cite{Hos12}.
\par
Quantum mechanics is built upon uncertainty relations between pairs of
canonical variables, of which the position and momentum of a particle
represent the prototype.
For example, in the ideal Heisenberg microscope, in which a photon of
energy $\hbar\,\omega$ is used to locate an electron, one finds the wavelength
of the photon sets a limit to the precision in, say, the position of the
electron along the $x$ coordinate given by~\footnote{Factors of order one
are neglected for simplicity.} 
\be
\Delta x
\gtrsim
\frac{1}{\omega\,\sin\theta}
\ ,
\ee
where $\theta$ represents the angular aperture of the microscope
within which the photon must travel in order to be seen.
At the same time, the electron will suffer a recoil
\be
\Delta p
\gtrsim
\hbar\,\omega\,\sin\theta
\ .
\ee
Putting the two bounds together one obtains the standard quantum
mechanical uncertainty 
\be
\Delta x\,\Delta p
\gtrsim
\hbar
\ ,
\ee
which suggests that infinite precision can be achieved in determining
$x$ at the price of giving up precision in $p$ (and {\em vice versa\/}).
Clearly, this requires photons scattering against the electron at higher and
higher energy.
\par
Now, a groundbreaking feature of gravity, which has not been proven
rigorously but shown to hold in great generality, is encoded by Thorne's
{\em hoop conjecture\/}~\cite{hoop}:
a black hole forms whenever the amount of energy $m$ is compacted 
inside a region that in no directions extends outside a circle of circumference
(roughly) equal to $2\,\pi\,\rh$, where
\be
\rh
=
\frac{2\,G\,m}{c^2}
=
2\,\lp\,\frac{m}{\mpl}
\ ,
\label{hoop}
\ee
is the gravitational Schwarzschild radius.
This result implies that once the energy $\hbar\,\omega$ has reached the
above threshold, instead of scattering off the electron and reach the 
microscope, the photon (along with the electron) will be trapped inside
a black hole and no measurement will occur.
Several arguments~\cite{Hos12} thus suggest the size of the
black hole contributes to the total uncertainty according to
\be
\Delta x
\gtrsim
\frac{\hbar}{\Delta p}
+
G\,\Delta p
\simeq
\lp
\left(
\frac{\mpl}{\Delta p}
+
\frac{\Delta p}{\mpl}
\right)
\ ,
\label{gup1}
\ee
where we estimated $m\sim\omega\sim\Delta p$.
It is then easy to see that Eq.~\eqref{gup1} leads to a minimum
uncertainty $\Delta x_{\rm min}\simeq\lp$ obtained for 
$\Delta p\simeq\mpl$.
\par
Let us also mention in passing that this kind of GUP can be formally derived
from modified canonical commutators and that it can be extended to the
models with extra spatial dimensions (see, {\em e.g.}, Ref.~\cite{SCgup})
of the type we shall consider in section~\ref{sec:extra}.
\subsection{GUP, horizon wave-function and particle collisions}
\label{ssec:hoopBH}
It is believed that black holes can form by the gravitational collapse of
astrophysical objects, such as the imploding cores of supernovae, or
by coalescing binary systems, which are the cases that originally inspired
the hoop conjecture.
Another possible mechanism is given by colliding particles, provided
the particles involved in the process have a sufficiently high energy and
small impact parameter to meet the requirements of the hoop conjecture.
Note that no minimum value of $m$ (or $\rh$) is however implied by this classical
conjecture.
\par
Once quantum physics is considered, black holes are expected to exist only above 
a minimum mass of the order of the fundamental scale of
gravity~\cite{Casadio:2013tma, Casadio:2013tza, Casadio:2013aua,Qhoop, Cminmass,Cminmass1,scale}.
In fact, if we neglect the spin and charge for the sake of simplicity,
general relativity associates to a point-like source of mass $m$ the gravitational
Schwarzschild radius~\eqref{hoop}, whereas quantum mechanics introduces
an uncertainty in the particle's spatial localisation, typically of the order of the
Compton length,
\be
\lambda_m
\simeq
\lp\,\frac{\mpl}{m}
\ .
\label{lambdaM}
\ee
Assuming quantum physics is a more refined description of classical physics,
the clash of the two lengths, $\rh$ and $\lambda_m$, implies that the former
only makes sense if it is larger than the latter, $\rh\gtrsim\lambda_m$.
In terms of the particle's mass, this means
\be
m
\gtrsim
\mpl
\ ,
\label{clM}
\ee
or the size of the black hole $\rh\gtrsim2\,\lp$.
Note that this bound is obtained from the flat space Compton length~\eqref{lambdaM},
but it is still reasonable to assume the condition~\eqref{clM} yields an
order of magnitude estimate of the minimum possible black hole mass.
Moreover, we have seen above that a minimum uncertainty of the same order
follows from GUPs (and modified commutators) precisely formulated in order
to account for black hole formation that should occur according to the hoop
conjecture.
\par
Instead of employing a GUP, we shall here show
that the above argument leading to Eq.~\eqref{clM} can be actually given a
probabilistic implementation, without modifying the commutators of quantum mechanics,
by associating to the particles an auxiliary
``horizon wave-function''~\cite{Casadio:2013tma, Casadio:2013tza, Casadio:2013aua}.
In order to introduce this tool, let us consider a state $\psi_{\rm S}$ representing
a massive particle localised in space and at rest in the chosen reference frame.
Having defined suitable Hamiltonian eigenmodes,
\be
\hat H\,\ket{\psi_E}=E\,\ket{\psi_E}
\ ,
\ee
where $H$ can be specified depending on the model we wish to consider,
the state $\psi_{\rm S}$ can be decomposed as
\be
\ket{\psi_{\rm S}}
=
\sum_E\,C(E)\,\ket{\psi_E}
\ .
\label{CE}
\ee
If we further assume the particle is spherically symmetric,
we can invert the expression of the Schwarzschild radius~\eqref{hoop}
to obtain $E$ as a function of $\rh$.
We then define the {\em horizon wave-function\/} as
\be
\psi_{\rm H}(\rh)
\propto
C\left(\mpl\,{\rh}/{2\,\lp}\right)
\ ,
\ee
whose normalisation can finally be fixed in a suitable inner product.
We interpret the normalised wave-function $\psi_{\rm H}$ as yielding the probability
that we would detect a horizon of areal radius $r=\rh$ associated with the particle in the
quantum state $\psi_{\rm S}$.
Such a horizon is necessarily ``fuzzy'', like the position of the particle itself,
and unlike its purely classical counterpart.
The probability density that the particle lies inside its own horizon of radius $r=\rh$
will next be given by
\be
P_<(r<\rh)
=
P_{\rm S}(r<\rh)\,P_{\rm H}(\rh)
\ ,
\label{PrlessH}
\ee
where
\be
P_{\rm S}(r<\rh)
=
4\,\pi\,\int_0^{\rh}
|\psi_{\rm S}(r)|^2\,r^2\,\d r
\ee
is the probability that the particle is found inside a sphere of radius $r=\rh$,
and
\be
P_{\rm H}(\rh)
=
4\,\pi\,\rh^2\,|\psi_{\rm H}(\rh)|^2
\ee
is the probability that the horizon is located on the sphere of radius $r=\rh$.
Finally, the probability that the particle described by the wave-function $\psi_{\rm S}$
is a black hole will be obtained by integrating~\eqref{PrlessH} over all possible
values of the radius,
\be
P_{\rm BH}
=
\int_0^\infty P_<(r<\rh)\,\d \rh
\ .
\label{PBH}
\ee
\par
\begin{figure}[t]
\centering
\raisebox{3.5cm}{$P_{\rm BH}$}
\includegraphics[width=7cm]{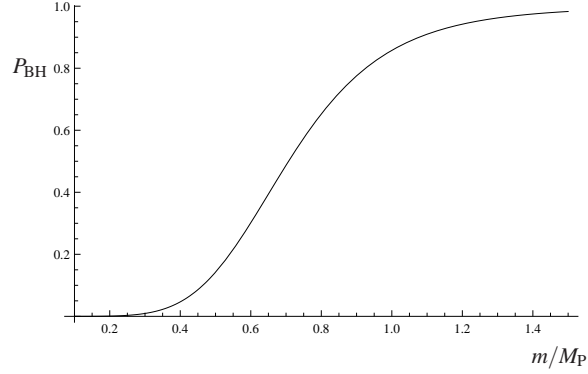}
\\
\hspace{7cm}$m/\mpl$
\caption{Probability $P_{\rm BH}$ that particle of mass $m$ is a black hole.
\label{Pbh}}
\end{figure}
\begin{figure}[t]
\centering
\raisebox{3.5cm}{$\frac{\Delta r}{\lp}$}
\includegraphics[width=8cm]{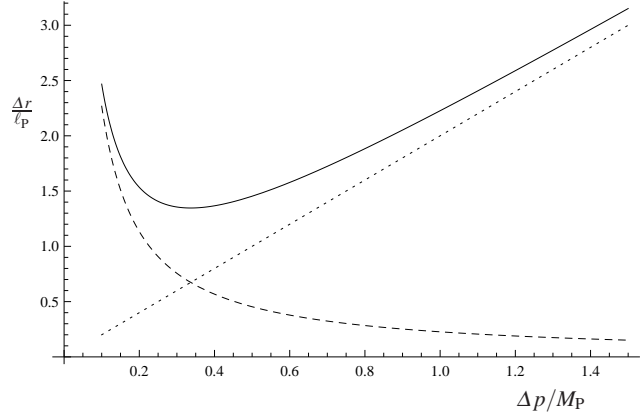}
\\
\hspace{6cm}$\Delta p/\mpl$
\caption{Uncertainty relation~\eqref{effGUP} (solid line) as a combination
of the quantum mechanical uncertainty (dashed line) and the uncertainty
in horizon radius (dotted line).
\label{pGUP}}
\end{figure}

As a check that this formalism leads to sensible results in agreement with the
bound~\eqref{clM}, one can easily apply it to a particle described by a
spherically symmetric Gaussian wave-function of width $\ell\simeq \hbar/m$,
\be
\psi_{\rm S}(r)
=
\frac{e^{-\frac{r^2}{2\,\ell^2}}}{\ell^{3/2}\,\pi^{3/4}}
\ ,
\ee
for which one obtains a vanishing probability that the particle is a black hole
when its mass is smaller than about
$\mpl/4$~\cite{Casadio:2013tma, Casadio:2013tza, Casadio:2013aua}
(see Fig.~\ref{Pbh}).
Moreover, by adding to the uncertainty in position $\Delta r$ determined by
the wave-function $\psi_{\rm S}$ the uncertainty in the size of the horizon
$\Delta \rh$ obtained from the corresponding horizon wave-function $\psi_{\rm H}$,
one is able to recover a GUP~\cite{Casadio:2013aua}.
In particular,
\be
\expec{\Delta r^2}
=
4\,\pi\,\int_0^{\infty}
|\psi_{\rm S}(r)|^2\,r^4\,\d r
\simeq
\ell^2
\ ,
\label{Dr}
\ee
and
\be
\expec{\Delta \rh^2}
=
4\,\pi\,\int_0^{\infty}
|\psi_{\rm H}(\rh)|^2\,\rh^4\,\d \rh
\simeq
\frac{\lp^4}{\ell^2}
\ .
\label{DRH}
\ee
Since
\be
\expec{\Delta p^2}
=
4\,\pi\,\int_0^{\infty}
|\psi_{\rm S}(p)|^2\,p^4\,\d p
\simeq
\mpl^2\,\frac{\lp^2}{\ell^2}
\equiv
\Delta p^2
\ ,
\ee
we can also write
\be
\ell^2
\simeq
\lp^2\,\frac{\mpl^2}{\Delta p^2}
\ .
\ee
Finally, by combining the uncertainty~\eqref{Dr} with \eqref{DRH} linearly,
we find
\be
\Delta r
\equiv
\sqrt{\expec{\Delta r^2}}
+
\gamma\,
\sqrt{\expec{\Delta \rh^2}}
\simeq
\lp\,\frac{\mpl}{\Delta p}
+
\gamma^2\,\lp\,\frac{\Delta p}{\mpl}
\ ,
\label{effGUP}
\ee
where $\gamma$ is a coefficient of order one, and the result is
plotted in Fig.~\ref{pGUP} (for $\gamma=1$).
This is precisely the kind of GUP leading to a minimum measurable length
\be
\Delta r
\gtrsim
\gamma\,\lp
\ ,
\ee
obtained for $\Delta p\simeq\mpl$.
\par
Of course, the physically interesting problem of particles colliding at very high energy,
and forming a black hole, is clearly going to require significantly more work and
overcoming much harder technical difficulties.
A flavour of what is going on can however be obtained by extending
the above construction to a state containing two free particles in one-dimensional
flat space~\cite{Qhoop}.
We represent each particle at a given time and position $X_i$ ($i=1$ or $2$)
by means of Gaussian wave-functions,
\be
\pro{x_i}{\psi_{\rm S}^{(i)}}
\equiv
\psi_{\rm S}(x_i)
=
e^{-i\,\frac{P_i\,x_i}{\hbar}}\,\frac{e^{-\frac{\left(x_i-X_i\right)^2}{2\,\ell_i}}}{\sqrt{\pi^{1/2}\,\ell_i}}
\ ,
\ee
where $\ell_i$ is the width and $P_i$ the linear momentum
(which remain constant).
The total wave-function is then just the product of the two one-particle states,
\be
\pro{x_1,x_2}{\psi_{\rm S}^{(1,2)}}
\equiv
\psi_{\rm S}(x_1,x_2)
=
\psi_{\rm S}(x_1)\,\psi_{\rm S}(x_2)
\ .
\label{PsiProd}
\ee
It is convenient to go through momentum space in order to compute the
spectral decomposition.
We find
\be
\pro{p_i}{\psi_{\rm S}^{(i)}}
\equiv
\psi_{\rm S}(p_i)
=
e^{-i\,\frac{p_i\,X_i}{\hbar}}\,
\frac{e^{-\frac{\left(p_i-P_i\right)^2}{2\,\Delta_i}}}{\sqrt{\pi^{1/2}\,\Delta_i}}
\ ,
\ee
where $\Delta_i=\hbar/\ell_i$, and we shall use the relativistic dispersion relation
\be
E_i
=
\sqrt{p_i^2+m_i^2}
\ .
\ee
If the particles were at rest ($P_i=0$), we could assume $\ell_i=\lambda_{m_i}$
(and $\Delta_i=m_i$).
For realistic elementary particles $m_1\simeq m_2\ll\mpl$, and, from Eq.~\eqref{clM}
 one expects
the probability of forming a black hole will become significant only
for $|P_i|\sim E_i\sim \mpl$.
From
$
P_i=\frac{m_i\,v_i}{\sqrt{1-v_i^2}}
$,
we obtain
\be
\ell_i
=
\frac{\hbar}{\sqrt{P_i^2+m_i^2}}
\simeq
\frac{\lp\,\mpl}{|P_i|}
\ ,
\qquad
{\rm and}
\qquad
\Delta_i
\simeq
|P_i|
\ .
\ee
The two-particle state can now be written as
\be
\ket{\psi_{\rm S}^{(1,2)}}
=
\prod_{i=1}^2
\left[
\int\limits_{-\infty}^{+\infty}
\d p_i
\,\psi_{\rm S}(p_i)\,\ket{p_i}
\right]
\ ,
\label{PsiPp}
\ee
and the relevant coefficients in the spectral decomposition~\eqref{CE} are given by
the sum of all the components of the product wave-function~\eqref{PsiPp}
corresponding to the same total energy $E$,
\be
C(E)
=
\!\!
\int\limits_{-\infty}^{+\infty}
\int\limits_{-\infty}^{+\infty}
\psi_{\rm S}(p_1)\,\psi_{\rm S}(p_2)\,
\delta(E-E_1-E_2)\,
\d p_1\,\d p_2
\ .
\label{C(E)}
\ee
For this two-particle collision, the horizon wave-function must be computed
in the centre-of-mass frame of the two-particle system,
so that
\be
P_1=-P_2
\equiv P>0
\ .
\ee
From $P\sim\mpl\gg m_1\simeq m_2$, we can also set
\be
X_1\simeq -X_2
\equiv 
X>0
\ .
\label{Xcm}
\ee
After replacing the expression of the Schwarzschild radius~\eqref{hoop}
into Eq.~\eqref{C(E)}, and (numerically) normalising the result in the inner product 
\be
\pro{\psi_{\rm H}}{\phi_{\rm H}}
\equiv
\int_0^\infty
\psi_{\rm H}^*(\rh)\,\phi_{\rm H}^*(\rh)\,\d\rh \ ,
\ee
we finally obtain the wave-function $\psi_{\rm H}=\psi_{\rm H}(\rh;X,P)$.
One then finds that $P_{\rm H}=|\psi_{\rm H}(\rh)|^2$ shows a mild dependence
on $X$ and a strong dependence on $P$,
in agreement with the fact that the energy of the system only depends on
$P$, and not on the spatial separation between the two particles.
It is also worth noting that $P_{\rm H}=P_{\rm H}(\rh)$ always peaks around
$\rh\simeq 2\,\lp\,(2 P/\mpl)$, in very good agreement with the
hoop conjecture~\eqref{hoop}.
\begin{figure}[t]
\centering
\includegraphics[width=9cm]{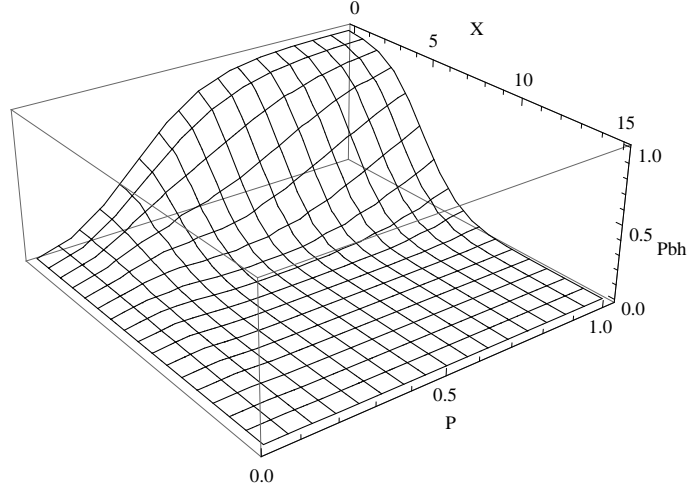}
\caption{Probability the two-particle system is a black hole as a function of $X$ and
$P$ (in units of Planck length and mass respectively).}
\label{P_BH}
\end{figure}
\par
The probability~\eqref{PBH} that the system of two particles is a black hole can next
be computed numerically as a function of the distance from the centre of mass $X$
of each particle, and the total energy $2 P$.
Fig.~\ref{P_BH} shows the result for a suitable range of $X$ and $P$.
Note that a first estimate of what happens as the two particles evolve in time can be
obtained by considering the probability $P_{\rm BH}=P_{\rm BH}(X,2P)$ along lines
of constant $P$ and decreasing $X$:
$P_{\rm BH}$ clearly increases up to the maximum reached for $X=0$,
when the two (non-interacting) particles exactly superpose.
There is therefore a significant probability that the collision produces a black hole,
say $P_{\rm BH}(X,2P\gtrsim 2\mpl)\gtrsim 80\%$, if the distance from the
centre of mass and linear momentum satisfy 
\be
X
\lesssim
{2\,\lp}\left({2P}/{\mpl}\right)-{\lp}
=
\rh(2P)-\lp
\ ,
\label{linearHoop}
\ee
where the term $-\lp$ on the right is the ``quantum mechanical correction'' to the hoop
formula~\eqref{hoop} for $E\simeq 2P\gtrsim 2\,\mpl$.
This correction is indeed arbitrary (as is the choice of $P_{\rm BH}\gtrsim 80\%$)
and applies to the formation of large (semi)classical black holes,
for which it is practically negligible.
For lower values of $P$, one instead finds 
$P_{\rm BH}(X,2P\lesssim 2\mpl)\gtrsim 80\%$ 
if
\be
2P-\mpl
\gtrsim
{\mpl\,X^2}/{9\,\lp^2}
\ ,
\label{quadHoop}
\ee
which yields the minimum energy $2P\simeq \mpl$ [instead of $2P\simeq \mpl/2$,
as it would follow from the linear relation~\eqref{linearHoop}].
Eq.~\eqref{quadHoop} represents a significant quantum mechanical correction
to the hoop conjecture~\eqref{hoop} for quantum black hole production,
that fully entails  the existence of a minimum black hole mass
(albeit a ``fuzzy'' one).
\par
Notably, the above result is obtained without assuming any specific microscopic structure
for the (quantum) black holes, and should therefore represent a quite generic 
expectation.
Similarly to the bound~\eqref{clM}, it implies that black holes fall well outside the
realm of experimental physics on earth.
There is however a catch, as we shall discuss in section~\ref{sec:extra}.
\subsection{Regular black holes}
\label{ssec:regBH}
We have just shown that (quantum) black holes cannot have arbitrarily small mass.
In this section we shall see that a similar conclusion also follows from considering
the possibility of improving the short distance behaviour of classical black hole metrics.
This is one of the key points for the validity of any candidate theory to quantum gravity.
However, despite the progress and the formulation of several approaches to quantum gravity,
the  mechanism of regularisation of black hole space-times still evades a complete
understanding (for instance, see Re.~\cite{BeR07}  for the case of string theory, 
and \cite{Rov08} for the case of loop quantum gravity). 

Given this background, one can address the problem of curvature singularities by
following alternative routes.
The earliest attempts of singularity avoidance were based on the assumption of a
de~Sitter core placed at the space-time origin.
Actually de~Sitter cores offer regular space-time regions where gravity becomes
locally repulsive and prevents a complete gravitational collapse into a singular
configuration.
Despite the effective nature of the approach, de~Sitter cores can be interpreted
as the effect of the graviton quantum vacuum energy.
This fact is confirmed by a local violation of energy conditions, which certify the
non-classical nature of the resulting space-time geometry. 

The first black hole with a regular de~Sitter core is maybe the Bardeen space-time~\cite{bardeen}.
This model has inspired a variety of additional improved black hole metrics,
based on different regularising mechanisms, such as: 
matching an outer Schwarzschild geometry and an inner de~Sitter geometry along
time-like~\cite{aa1,aa2,aa3} and space-like~\cite{fmm} matter  shells;
coupling of gravity with non-linear electrodynamics~\cite{beato};
prescribing a stress tensor for the quantum vacuum energy density at the
origin~\cite{dym,dym2,MbK05,MbK08};
implementing quantum gravity effects in classical
backgrounds~\cite{hay,SpS12,NiS14,Mod06,MoP09,BoR00,Car13,CMP11}
(for a review see Ref.~\cite{stefano}).
Despite the progress and virtues of such proposals, one finds that the resulting
metrics are affected by either one or a combination of the following factors:
i) lack of a stress tensor;
ii) lack of a natural transition between inner and asymptotic geometries;
iii) lack of a neutral solution;
iv) need of additional hypotheses to achieve the regularity;
v) lack of equations describing the space-time geometry at all distances;
vi) lack of a clear connection with some quantum gravity formulation. 
 
A simple way to overcome the above limitations is offered by the so-called
non-commutative geometry inspired black hole solutions~\cite{Nic09}.
Here, we briefly summarise the procedure to derive such a family of regular black holes.
One can start by considering the action $S_{\mathrm{tot}}$ describing the
space-time generated by a static massive source
\begin{equation}
S_{\mathrm{tot}}=S_{\mathrm{grav}}+S_{\mathrm{matt}}~,
\label{totaction}
\end{equation}
where $S_{\mathrm{grav}}$ is the usual Einstein-Hilbert action, while the
matter action $S_{\mathrm{matt}}$ reads
\begin{equation}
S_{\mathrm{matt}}=
-\int \,  \mathrm{d}^{4} x \, \sqrt{-g} \ \rho(x),\ \quad \rho(x)
=\frac{M}{\sqrt{-g}} \int \,  \mathrm{d}\tau  \ \delta\left(x-x(\tau)\right)
\ ,
\end{equation}
with $\rho(x)$ being the energy density describing a massive, point-like particle.
By varying (\ref{totaction}) with respect to the metric $g_{\mu\nu}$, one finds
the Einstein's equations with a pressure-less source term 
\begin{eqnarray}
 T^0\,_0=-\frac{M}{4\pi\, r^2}\delta ( r )
 \ .
 \label{t00}
\end{eqnarray}
Customarily, one ignores the distributional profile of the source, \textit{i.e.}, $\delta ( r )$,
accounting for the singular behaviour of the resulting space-time geometry.
Rather one prefers to solve the Einstein equations outside the source, \textit{i.e.},
in an open domain ${\cal D}=\mathbb{R}^4\setminus\vec{0}$.
The Schwarzschild solution is then obtained by integrating Einstein's equations
\textit{\`a la} Laplace, namely by exploiting boundary conditions.
Although mathematically acceptable, this procedure has several drawbacks
from a physical viewpoint:
for instance, the mass term emerges as an integration constant and it is placed
at the point that one has excluded by hypothesis.
Not surprisingly black hole solutions are often labeled as \textit{vacuum solutions},
a definition that might sound  in contradiction to the basic
tenet of general relativity according to which gravity, expressed in terms of curvature,
is the result of the presence of mass and energy in the space-time
(for additional details see Ref.~\cite{BaN93,BaN94}).

More importantly, Eq.~\eqref{t00} is instrumental for the present discussion:
we expect any candidate theory of quantum gravity to improve the source term,
by providing a new, ultraviolet finite profile for the energy density.
This is the case of non-commutative geometry, which is based on the idea of
implementing a fundamental length by allowing a non-trivial commutation relation
for coordinate operators
\begin{equation}
\left[X^\mu, X^\nu\right]=i\,\Theta^{\mu\nu}
\ ,
\label{ncg}
\end{equation}
where $\Theta^{\mu\nu}$ is a constant anti-symmetric tensor with determinant
$|\Theta^{\mu\nu}|=\theta$, having units of a length squared.
The parameter $\sqrt{\theta}$ will act as a quantum of length and it is natural
to assume $\sqrt{\theta}=\ell_\mathrm{P}$.
However $\sqrt{\theta}$ is not fixed \textit{a priori} and can be treated as a
parameter adjustable to those scales at which non-commutative effects set in.
By averaging coordinate operators $X^\mu$ on suitable coherent states,
it has been shown, in a series of papers~\cite{SmS03a,SmS03b,SmS04,SSN06,KoN10, Casadio:2005vg, Casadio:2007ec},
that the integration measures in momentum space, are exponentially suppressed
in the ultraviolet sector by a factor $\exp(-\theta k^2)$, where  $k$ is the magnitude
of the Euclidean momentum.
As a result, by adopting free falling Cartesian-like coordinates, we have that the
usual representation of the source term \eqref{t00} switches to a new, regular profile 
\begin{equation}
\delta^{(3)}(\vec{x})
=
\frac{1}{(2\pi)^3}\int  \mathrm{d}^3 k\, e^{i\vec{k}\cdot\vec{x}}
\rightarrow
\rho_\theta(\vec{x})
=
\frac{1}{(2\pi)^3}\int  \mathrm{d}^3 k\, e^{-\theta k^2+ i\vec{k}\cdot\vec{x}}
=
\frac{e^{-{x}^2/4\theta}}{(4\pi \theta)^{3/2}}
\ ,
\label{gauss}
\end{equation}
namely a Gaussian distribution whose width is the minimal resolution length
$\sqrt{\theta}$ \cite{NSS05,Nic05}.
This result consistently reproduces the classical limit \eqref{t00} for
$\sqrt{\theta}\to 0$.
More importantly the Gaussian approaches a finite, constant value at the
origin as expected for a de~Sitter core.
The latter requires negative pressure terms to sustain the Gaussian profile
and prevent the collapse into a Dirac delta distribution.
By the conservation of the energy-momentum tensor $\nabla_\mu T^{\mu\nu}=0$,
one can determine all pressure terms, namely the radial pressure
$p_\mathrm{rad}(r)=-M\rho_\theta(r)$ and the angular pressure
$p_\perp(r)=-M\rho_\theta(r)-\frac{r}{2}\ M\partial_r \rho_\theta(r)$,
where
\be
{T}_\nu^{\ \mu}=\mathrm{diag}\left(-M\rho_\theta(r), p_\mathrm{rad}(r), p_\perp(r), p_\perp(r)\right)
\ .
\ee
As a result the energy momentum tensor describes an anisotropic fluid,
that violates energy conditions in the vicinity of the origin, as expected. 

By solving the Einstein equations with the above source term one finds
\begin{eqnarray}
\mathrm{d}s^2
&\!\!=\!\!&
-\left(1-\frac{2MG}{r}\frac{\gamma(3/2; r^2/4\theta)}{\Gamma(3/2)}\right)~\mathrm{d}t^2 + \left(1-\frac{2MG}{r}\frac{\gamma(3/2; r^2/4\theta)}{\Gamma(3/2)}\right)^{-1}~\mathrm{d}r^2
\nonumber
\\
&&
+r^2\,\mathrm{d}\Omega^2 
\label{ncschw}
\end{eqnarray}
known as the non-commutative geometry inspired Schwarzschild solution~\cite{NSS06}.
Here $\gamma(a/b,x)=\int_0^x \frac{\mathrm{d}t}{t}t^{a/b}e^{-t}$ is the incomplete Euler
gamma function and $\Gamma(3/2)=\frac{1}{2}\sqrt{\pi}$.
Before presenting the properties of the line element \eqref{ncschw}, we recall that the
above results has been confirmed by two alternative derivations:
the Gaussian profile~\eqref{gauss} can be obtained by means of a Voros product
approach to non-commutative geometry~\cite{BGM10}, as well as by solving the
gravitational field equations derived by non-local deformations of Einstein gravity~\cite{MMN11}.
In the latter case, the non-local character of gravity is an alternative way to accounting
for the presence of a minimal resolution length at extreme energy scales
(see~\cite{Mof11,Nic12,IMN13,Mod12,BGK12} for recent non-local gravity proposals). 
Interestingly the space-time~\eqref{ncschw} has been interpreted as a condensate of
gravitons~\cite{CaO13}, according to a recently proposed conjecture about the nature of
black holes~\cite{DvG11,DvG13a,DvG13b,DFG13}.

The line element~\eqref{ncschw} consistently matches the Schwarzschild geometry in the large 
distance limit $r\gg\sqrt{\theta}$.
Conversely at small distance, $r\sim\sqrt{\theta}$, the 
incomplete Euler gamma function goes like $\gamma(3/2,r^2/4\theta)\sim r^3$ and therefore a 
deSitter core develops at the origin.
The curvature is easily obtained:
the Ricci scalar is finite, constant and positive at the origin and reads 
\be
R(0)=\frac{4M}{\sqrt{\pi}\theta^{3/2}}
\ .
\ee
\begin{figure}[t]
\centering
\includegraphics[width=8cm]{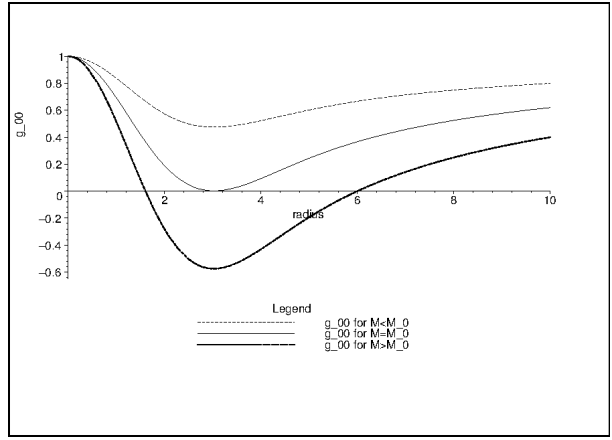}
\caption{The metric element $g_{00}=1-\frac{2MG}{r}\frac{\gamma(3/2; r^2/4\theta)}{\Gamma(3/2)}$
as a function of $r$ for different values of the parameter $M$.}
\label{g00Schw}
\end{figure}
\par
The analysis of the horizon equation requires a preliminary comment.
Here the parameter $M$ is defined as the integrated flux of energy
\begin{equation}
M\equiv \int_\Sigma \mathrm{d}\sigma^\mu\, T_\mu^0
\end{equation}
where $\Sigma$ is an asymptotic closed space-like surface.
Equivalently $M$  results as the limit $r\to\infty$ of the cumulative mass distribution
\begin{equation}
m(r) = - 4\pi \int_0^r \d t\, t^2\, T^0\,_0 
\ .
\end{equation}
Depending on the values of $M$, the horizon equation
$1-\frac{2MG}{r}\frac{\gamma(3/2; r^2/4\theta)}{\Gamma(3/2)}=0$ has two, one or no solutions.
Specifically there exists a value $M_0\simeq 1.90 \,\sqrt{\theta}/G$ for the mass parameter
such that (see Fig. \ref{g00Schw})
\begin{itemize}
\item for $M<M_0$ there is no solution, corresponding to the case of a regular manifold without horizons;
\item for $M>M_0$ there are two solutions, corresponding to an inner horizon $r_-$ and an outer horizon $r_+$:
\item for $M=M_0$ there is just one solution, corresponding to a single degenerate
horizon $r_0\simeq 3.02 \sqrt{\theta}$.
\end{itemize}
The global structure of the solution resembles that of the Reissner-Norstr\"{o}m geometry,
with a never ending chain of space-time quadrants.
However, contrary to the latter case, there is no singularity.
The global structure of the solution differs also from that of the Kerr geometry
because the space-time is geodesically complete.
As a result the negative $r$ geometry does not represent an analytical continuation
of Eq.~\eqref{ncschw}, but rather an additional, horizonless, regular space-time.
Horizons have the conventional meaning, \textit{i.e.}, $r_+$ is an event horizon and
$r_-$ is a Cauchy horizon.
Analyses of possible blue shift instability at $r_-$ are controversial~\cite{BaN10,BrM11}:
it is not yet clear how quantum gravity effects might tame the occurrence
of a possible mass inflation.
Finally $r_0$ is the radius of the extremal black hole, we expect it to be a zero
temperature configuration. 
\begin{figure}[t]
\centering
\includegraphics[width=8cm]{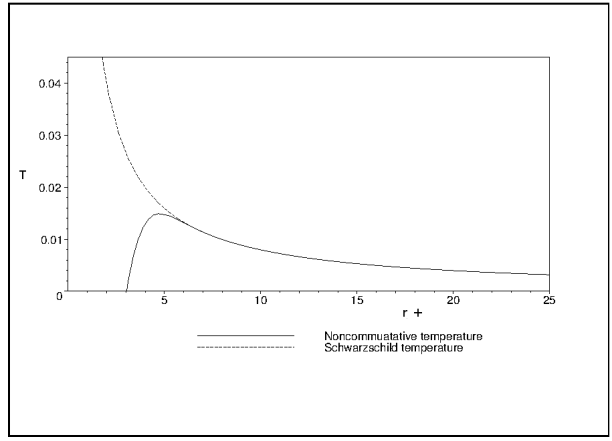}
\caption{The metric element $g_{00}=1-\frac{2MG}{r}\frac{\gamma(3/2; r^2/4\theta)}{\Gamma(3/2)}$
as a function of $r$ for different values of the parameter $M$.}
\label{temp}
\end{figure}
\par
The thermodynamics of the non-commutative geometry inspired black hole~\cite{BMS08}
can be studied as follows.
The periodicity of the imaginary time of the metric~\eqref{ncschw} gives
the temperature of the black hole (see Fig.~\ref{temp})
\begin{equation}
T
=
\frac{1}{4\pi r_+}\left(1-\frac{r_+^3}{4\theta^{3.2}}\frac{e^{-r_+^2/4\theta}}{\gamma(3/2; r_+^2/4\theta)}\right)
\ .
\label{temperature}
\end{equation}
We see that at large distances $r_+\gg\sqrt{\theta}$ the non-commutative
corrections are exponentially vanishing and the temperature matches the
Hawking result $\sim 1/r_+$.
On the other hand, at shorter distances, \textit{i.e.}, at $r_+\simeq 4.76\sqrt{\theta}$
the temperature admits a maximum $T_\mathrm{max}\simeq 0.015 \,\sqrt{\theta}/G$,
signalling the presence of a transition from a negative heat capacity phase
to a positive heat capacity phase.
The final stage of the evaporation is completely new:
instead of the runaway divergent behaviour of the temperature,
the black hole slowly cools down towards the zero temperature extremal
configuration at $r_0$.
Such a new terminal phase of the evaporation, often called ``SCRAM phase''~\footnote{One borrows
the terminology of critical shutdowns of thermonuclear reactors}
downplays any concerns about the quantum back reaction.
One can check that the emitted energy is always negligible with respect to
black hole mass, \textit{i.e.}, $T/M<T_\mathrm{max}/M_0\simeq 7.89\times 10^{-3}$.
This is equivalent to saying that the metric correctly describes the system gravity/matter
during all the evaporation process.

The presence of a phase transition from an unstable phase to a locally stable
SCRAM can be seen by analysing the heat capacity of the hole.
As a preliminary step one calculates the black hole entropy
by integrating $\mathrm{d}S\equiv \mathrm{d}M/T$, which yields
\begin{eqnarray}
S
=
\frac{1}{4}\left(\frac{A_+\ \Gamma(3/2)}{\gamma(3/2; r_+^2/4\theta)}-\frac{A_0\ \Gamma(3/2)}{\gamma(3/2; r_0^2/4\theta)}\right)+\frac{\pi}{2\theta^{3/2}}\int_{r_0}^{r_+}\frac{t^2e^{-t^2/4\theta}dt}{\gamma^2(3/2; t^2/4\theta)}
\ ,
\label{Entropy}
\end{eqnarray}
where $A_+=4\pi r_+^2$ and $A_0=4\pi r_0^2$.
We see that, for large holes, \textit{i.e.}, $A_+\gg A_0$, Eq.~\eqref{Entropy}
reproduces the usual area law.
On the other hand for smaller holes, the presence of the extremal configuration
becomes important and leads to a vanishing entropy for $r_+=r_0$.
This fact implies that the extremal configuration is a stable remnant of the evaporation.
The stability of the remnant can also been seen through the  black hole heat capacity  
\begin{equation}
C(r_+)=T_H \left(\frac{dS_H}{dr_+}\right)\left(\frac{dT_H}{dr_+}\right)^{-1}
\ .
\end{equation}
which vanishes at $r_0$.
As a result remnants are extremal black hole configurations with $T=S=C=0$.
More importantly, $C$ admits an asymptote  $\frac{dT_H}{dr_+}=0$, \textit{i.e.},
at $T_\mathrm{max}$ which corresponds to a transition from a un-stable to a
stable phase preceding the remnant formation.
Such properties greatly improve the scenario based on the GUP~\cite{ACS01,Sca99,ChA03},
which suffers from the following weak points:
huge back reaction due to Planckian values of remnant temperature;
instability due a negative heat capacity in the phase preceding the remnant formation;
sign ambiguity in the expression of the temperature;
absence of any metric whose surface gravity reproduces the black hole temperature.

For the above attracting feature, the metric~\eqref{ncschw} has been studied in several
contexts.
For instance it has been shown that zero temperature remnants might have copiously
been produces during the early ages of the Universe, as a consequence of the de~Sitter space
quantum instability~\cite{MaN11}.
On the other hand, the novel thermodynamic properties have been displayed by
considering an anti-de~Sitter background for~\eqref{ncschw}:
the intriguing new feature is the possibility of improving the conventional Hawking-Page
phase transition in terms of a real gas phase diagram.
In the isomorphism of variables, the black hole remnant size actually plays the role of
the constant $b^\prime $ representing the molecule size in the van der~Waals
theory~\cite{NiT11,SmS13,SpS13}.
 
The metric~\eqref{ncschw} has companion geometries like traversable wormholes~\cite{GaL09},
whose throat is sustained by negative pressure terms, dirty black holes~\cite{NiS10} and
collapsing matter shells~\cite{NOS13}.
More importantly, the non-commutative geometry inspired Schwarzschild black hole
has been studied  in the presence of large extra dimensions~\cite{Riz06}.
As a special result, higher-dimensional non-commutative black holes
tend to emit softer particles mainly on the brane, in marked contrast with
the emission spectra of conventional Schwarzschild-Tangherlini
black holes~\cite{CaN08,Gin10,NiW11}.
This peculiar emission spectrum might be a distinctive signature for detecting
black holes resulting from particle collisions.
However, the energy required for black hole formation might exceed current
accelerator capabilities, as explained in Refs.~\cite{MNS12,BlN14}.
Lower dimensional versions of the metric~\eqref{ncschw} have also been studied
in the context of dilatonic gravity:
surprisingly, the regularity of the manifold gives rise to a richer topology,
admitting up to six horizons~\cite{MuN11}.

Finally non-commutativity inspired black holes have been extended by including
all possible black hole parameters.
Charged~\cite{ANS07,SSN09}, rotating~\cite{SmS10}, and charged rotating~\cite{MoN10}
black holes have been derived in order to improve the Reissner-Nordstr\"{o}m,
Kerr and Kerr-Newman geometries.
Specifically in the case of rotating black holes, the cure of the ring singularit
is accompanied by the absence of conventional pathologies of the Kerr metric,
such as  an ``anti-gravity'' universe with causality violating time-like closed
world-lines and a ``super-luminal'' matter disk.
\section{Extra dimensions}
\label{sec:extra}
Models of the Universe with large additional dimensions were proposed around the
year 2000 to bypass the constraints of not having observable Kaluza-Klein modes.
In these scenarios the Standard Model particles and interactions are confined
on a thin ``brane'' embedded in a higher-dimensional space-time, while gravity leaks
into the extra dimensions~\cite{ArkaniHamed:1998rs, ArkaniHamed:1998nn, Antoniadis:1998ig, Randall:1999vf, Randall:1999ee}.
Because gravity propagates in the entire ``bulk" space-time, its fundamental
scale $\mg$ is related to the observed Planck mass $\mpl\simeq 10^{16}\,$TeV by a
coefficient determined by the volume of the (large or warped) extra dimensions.
Therefore in these models there appear several length scales, namely the spatial
extension(s) $L$ of the extra dimensions in the ADD scenario~\cite{ArkaniHamed:1998rs, ArkaniHamed:1998nn, Antoniadis:1998ig}, or
the anti-de~Sitter scale in the RS scenario $\ell$~\cite{Randall:1999vf, Randall:1999ee}, and possibly the finite
thickness $\Delta$ of the brane in either.
The size $L$ of the extra dimensions or the scale $\ell$, determines the value of the
effective Planck mass $\mpl$ from the fundamental gravitational mass $\mg$.
At the same time all of them determine the scale below which one should measure
significant departures from the Newton law.
\par
For suitable choices of $L$ or $\ell$, and the number $d$ of extra dimensions,
the mass $\mg$ in these scenarios can be anywhere below $\mpl\simeq 10^{16}\,$TeV,
even as low as the electro-weak scale, that is $\mg\simeq 1\,$TeV.
This means that the scale of gravity may be within the experimental reach of our
high-energy laboratories or at least in the range of energies of
ultra-high energy cosmic rays.
\subsection{Black holes in extra dimensions}
\label{ssec:extraBH}
%
%
We showed how we expect black holes can exist only above a minimum mass
of the order of the fundamental scale of gravity.
In four dimensions, this value is about $10^{16}\,$TeV, and it would therefore be
impossible to produce black holes at particle colliders
or via the interactions between ultra-high cosmic rays with nucleons in
the atmosphere. 
However, if we live in a universe with more than three spatial dimensions,
microscopic black holes with masses of the order of $\mg\simeq 1\,$TeV may
be produced by colliding particles in present accelerators or by ultra-high
cosmic rays or neutrinos  (see, {\em e.g.\/}, Refs.~\cite{ Cavaglia:2002si, Kanti:2004nr, Cardoso:2012qm, Park:2012fe, Calmet:2012mf, Arsene:2013nca, Arsene:2013ria}).
\par
Our understanding of these scattering processes in models with extra spatial
dimensions now goes beyond the naive hoop conjecture~\cite{hoop} used
in the first papers on the topic.
After the black hole is formed, all of its ``hair'' will be released 
in the subsequent balding phase. 
If the mass is still sufficiently large, the Hawking radiation~\cite{hawking} will set off.
The standard description of this famous effect is based on the canonical Planckian distribution
for the emitted particles, which implies the life-time of microscopic black holes is very short,
of the order of $10^{-26}\,$s~\cite{dimopoulos, Banks:1999gd, Giddings:2001bu}.
This picture (mostly restricted to the ADD scenario~\cite{ArkaniHamed:1998rs, ArkaniHamed:1998nn, Antoniadis:1998ig}) has been implemented
in several numerical codes~\cite{charybdis,trunoir,ahn,catfish,cha2, Alberghi:2006qr,blackmax,charybdis2, sampaio},
mainly designed to help us identify black hole events at the Large Hadron Collider (LHC).
\par
We should emphasise that the end-stage of the black hole evaporation remains
an open problem to date~\cite{Landsberg:2006mm,Harris:2004xt,Casanova:2005id},
because we do not yet have a confirmed theory of quantum gravity.
The semiclassical Hawking temperature grows without bound,
as the black hole mass approaches the Planck mass. This is a sign of the lack of predictability
of perturbative approaches, in which the effect of the Hawking radiation on the
evaporating black hole is assumed to proceed adiabatically (very slowly).
Alternatively one can use the more consistent microcanonical description
of black hole evaporation, in which energy conservation is granted by
construction~\cite{mfd, mfd1, mfd2, entropy}.
This would seem an important issue also on the experimental side, since the
microcanonical description predicts deviations from the Hawking law for small
black hole masses (near the fundamental scale $\mg$) and could lead to
detectable signatures.
However, energy conservation is always enforced in the numerical codes,
and the deviations from the standard Hawking formulation are thus masked
when the black hole mass approaches $\mg$~\cite{cha2, Alberghi:2006qr}.
The default option for the end-point of microscopic black holes in most codes
is that they are set to decay into a few standard model particles when a 
low mass (of choice) is reached. 
Another possibility, with qualitatively different phenomenology, is for the evaporation
to end by leaving a stable remnant of mass
$M\simeq \mg$~\cite{Koch:2005ks,Hossenfelder:2005bd,gingrich}.
\par
Given the recent lower bounds on the value of the Planck mass,
it has been pointed out that semiclassical black holes seem to be difficult
to produce at colliders, as they might indeed require energies 5 to 20 times
larger than the Planck scale $\mg$.
Similar objects, that generically go under the name of ``quantum black holes'',
could be copiously produced
instead~\cite{Calmet:2008dg,Calmet:2011ta,Calmet:2012cn,Calmet:2012mf}.
Their precise definition is not settled, but one usually assumes their production
cross section is the same as that of larger black holes, and they are non-thermal objects,
which do not decay according to the Hawking formula.
Their masses are close to the scale $\mg$ and their decay might resemble strong gravitational
rescattering events~\cite{Meade:2007sz}. 
It is also typically assumed that non-thermal quantum black holes decay into only
a couple of particles.
However, depending on the details of quantum gravity, the smallest quantum black holes
could also be stable and not decay at all.
The existence of remnants, i.e.~the smallest stable black holes, have been considered
in the literature~\cite{Koch:2005ks,Hossenfelder:2005bd}, and most of the results
presented here can be found in Refs.~\cite{Bellagamba:2012wz,Alberghi:2013hca}.
\subsubsection{Black Hole Production}
\label{production}
In the absence of a quantum theory of gravity, the production cross section
of quantum black holes is usually extrapolated from the semiclassical regime.
Therefore, both semiclassical and quantum black holes are produced
according to the geometrical cross section formula extrapolated from the
(classical) hoop conjecture~\cite{hoop}, 
\be
\sigma_{\rm BH}(M)\approx \pi\,R_{\rm H}^2
\ ,
\ee
and is thus proportional to the horizon area.
The specific coefficient of proportionality depends on the details of the models,
which is assumed of order one. 
\par 
In higher-dimensional theories, the horizon radius depends on the number
$d$ of extra-dimensions,
\be
R_{\rm H}
=
\frac{\ell_{\rm G}}{\sqrt{\pi}}\,
\left(\frac{M}{M_{\rm G}}\right)^{\frac{1}{d+1}}
\left(\frac{8\,\Gamma\left(\frac{d+3}{2}\right)}{d+2}
\right)^{\frac{1}{d+1}}
\ ,
\ee
where $\ell_{\rm G}=\hbar/\mg$ is the fundamental gravitational
length associated with $\mg$,
$M$ the black hole mass, $\Gamma$ the Gamma function,
and the four-dimensional Schwarzschild radius~\eqref{hoop}
is recovered for $d=0$.
The Hawking temperature associated with the horizon is thus
\be
T_{\rm H} =
\frac{d+1}{4\,\pi\,R_{\rm H}}
\ .
\label{TH}
\ee
In a hadron collider like the LHC, a black hole could form in the collision
of two partons, i.e.~the quarks, anti-quarks and gluons of the colliding protons
$p$.
The total cross section for a process leaving a black hole and other
products (collectively denoted by $X$) is thus given by
\be
\left.\frac{d\sigma}{d M}\right|_{pp\to {\rm BH}+X}
=\frac{dL}{d M}\,\sigma_{\rm BH}(ab\to BH; \hat s=M^2)
\ ,
\ee
where 
\be
\frac{dL}{d M}=\frac{2\,M}{s}\,\sum_{a,b}\int_{M^2/s}^1
\frac{dx_a}{x_a}\,f_a(x_a)\,f_b\left(\frac{M^2}{s\,x_a}\right)
\ ,
\ee
$a$ and $b$ represent the partons which form the black hole,
$\sqrt{\hat s}$ is their centre-mass energy, $f_i(x_i)$ are
parton distribution functions (PDF), and $\sqrt{s}$ is the centre-mass
collision energy.
We recall that the LHC data is currently available at $\sqrt{s}=8\,$TeV,
with a planned maximum of $14\,$TeV.
\subsubsection{Charged Black Holes}
\label{BWBH}
It is important to note that, since black holes could be produced via the
interaction of electrically charged partons (the quarks), they could carry
a non vanishing electric charge, although the charge might be preferably
emitted in a very short time. 
In four dimensions, where the fundamental scale of gravity is the Planck
mass $\mpl$, the electron charge $e$ is sufficient to turn such small objects
into naked singularities.
This can also be shown to hold in models with extra-spatial dimensions
for black holes with mass $M\sim \mg$ and charge $Q\sim e$.
However, since the brane self-gravity is not neglected in brane-world models
of the RS scenario~\cite{Randall:1999vf, Randall:1999ee}, a matter source located on the brane will give
rise to a modified energy momentum tensor in the Einstein equations projected
on the three-brane~\cite{shiromizu}.
By solving the latter, one finds that this backreaction can be described
in the form of a tidal ``charge'' $q$ which can take both positive and negative
values~\cite{dadhich}.
The interesting range of values for $q$ are the positive ones. 
Provided the tidal charge is large enough, microscopic
black holes can now carry an electric charge of the order of $e$~\cite{CH}.
In this particular case, the horizon radius is given by
\be
R_{\rm H}
=
\ell_{\rm P}\,
\frac{M}{M_{\rm P}}
\left(
1+
\sqrt{1-\tilde Q^2\,\frac{M_{\rm P}^2}{M^2}
+\frac{q\,M_{\rm P}^2}{\ell_{\rm G}^2\,M^2}}
\right)
\ ,
\label{tidalH}
\ee
where $\tilde Q$ is the electric charge in dimensionless units,
\be
\tilde Q
\simeq
10^8\left(\frac{M}{M_{\rm P}}\right)
\frac{Q}{e}
\ .
\ee
Reality of Eq.~\eqref{tidalH} for a remnant of charge $Q=\pm e$
and mass $M\simeq M_{\rm G}$ then requires
\be
q
\gtrsim
10^{16}\,\ell_{\rm G}^2
\left(\frac{M_{\rm G}}{M_{\rm P}}\right)^2
\sim
10^{-16}\,\ell_{\rm G}^2
\ .
\ee
Configurations satisfying the above bound were indeed found recently~\cite{covalle, covalle1}.
\subsubsection{Black Hole Evolution}
In the standard picture, the evolution and decay process of semiclassical
black holes can be divided into three characteristic stages:
\begin{enumerate}
\item
{\bf Balding phase.}
Since no collision is perfectly axially symmetric, the initial state will not be described by
a Kerr-Newman metric.
Because of the no-hair theorems, the black hole will therefore radiate away
the multipole moments inherited from the initial configuration, and reach a hairless state.
A fraction of the initial mass will also be radiated as gravitational radiation, on the brane
and into the bulk.
\item
{\bf Evaporation phase.}
The black hole loses mass via the Hawking effect.
It first spins down by emitting the initial angular momentum,
after which it proceeds with the emission of thermally distributed
quanta.
The radiation spectrum contains all the standard model particles,
(emitted on our brane), as well as gravitons (also emitted into the
extra~dimensions).
For this stage, it is crucial to have a good estimate of the grey-body
factors~\cite{ida, ida1, ida2,kanti, kanti1, kanti2, kanti3, kanti4}.
\item
{\bf Planck phase.}
\label{Pf}
The black hole has reached a mass close to the effective Planck
scale $M_{\rm G}$ and falls into the regime of quantum gravity.
It is generally assumed that the black hole will either completely
decay into standard model particles~\cite{dimopoulos, Banks:1999gd, Giddings:2001bu}
or a (meta-)stable remnant is left, which carries away the remaining energy~\cite{Koch:2005ks}.
\end{enumerate}
Admittedly, we have limited theoretical knowledge of the nature of quantum
black holes, since these objects should be produced already at stage~\ref{Pf}.
We should therefore keep our analysis open to all possible qualitative behaviours.
In particular, we shall focus on the case in which the initial semiclassical or
quantum black hole emits at most a fraction of its mass in a few particles and
lives long enough to exit the detector.
In other words, we will here focus on the possibility that the third phase ends
by leaving a (sufficiently) stable remnant.
\subsection{Minimum mass and remnant phenomenology }
\label{ssec:remnants}
It was shown in a series of articles \cite{Casadio:2009sz, Casadio:2010dq}  that it is possible
for Planck scale black holes to result in stable remnants.
Given the present lower bounds on the value of the fundamental scale $\mg$,
the centre of mass energy of the LHC is only large enough to produce
quantum black holes.
(We remind the mass of the lightest semiclassical black holes is expected
to be between 5 and 20 times $\mg$, depending on the model.)
In this case, the remnant black holes could not be the end-point of the
Hawking evaporation, but should be produced directly.
\par
At the LHC, black holes could be produced by quarks, anti-quarks and
gluons, and would thus typically carry a SU(3)$_c$ charge
(as well as a QED charge, as we pointed out before).
Quantum black holes could in fact be classified according to representations of SU(3)$_c$,
and their masses are also expected to be quantised~\cite{Calmet:2012cn}.
Since we are considering the case that black holes do not decay completely,
we expect that they will hadronize, i.e.~absorbe a particle charged under $SU(3)_c$
after traveling over a distance of some $200^{-1}\,$MeV and become an $SU(3)_c$ singlet.
They could also loose colour charge by emitting a fraction of their
energy before becoming stable.
Finally, the hadronization process could possibly lead to remnants with a (fractional) QED
charge.
To summarise, black hole remnants could be neutral or have the following QED charges:
$\pm 4/3$, $\pm 1$, $\pm 2/3$, and $\pm 1/3$.
Depending on its momentum, a fast moving black hole is likely to hadronize in the detector,
whereas for a black hole which is moving slowly, this is likely to happen before it reaches
the detector.
\par
Monte Carlo simulations of black hole production processes which result in stable remnants
have been performed using the code CHARYBDIS2. 
They have shown that approximately $10\%$ of the remnants will carry an electric charge
$Q=\pm e$~\cite{Bellagamba:2012wz}.
This code was not specifically designed to simulate the phenomenology
of quantum black holes, but it could be employed since they are produced
according to the same geometrical cross section as semiclassical black holes,
and the details of their possible partial decay are not phenomenologically relevant
when searching for a signature of the existence of remnants.
In fact, the initial black hole mass cannot be much larger than a few times $M_{\rm G}$,
even for $\sqrt{s}=14\,$TeV.
So in the simulations the black holes emit at most a fraction of their energy in a
small number of standard model particles before becoming stable.
Such a discrete emission process in a relatively narrow range of masses
is constrained by the conservation of energy and standard model charges,
and cannot differ significantly for different couplings of the quantum black holes
to standard model particles.
In Monte Carlo generators the decays are assumed to be instantaneous.
The following analysis does therefore not include the possibility that the black holes
partially decay off the production vertex, nor the effects of hadronization by absorption
of coloured particles.
\begin{figure}[t]
\centerline{\begin{tabular}{ cc}
\epsfig{figure=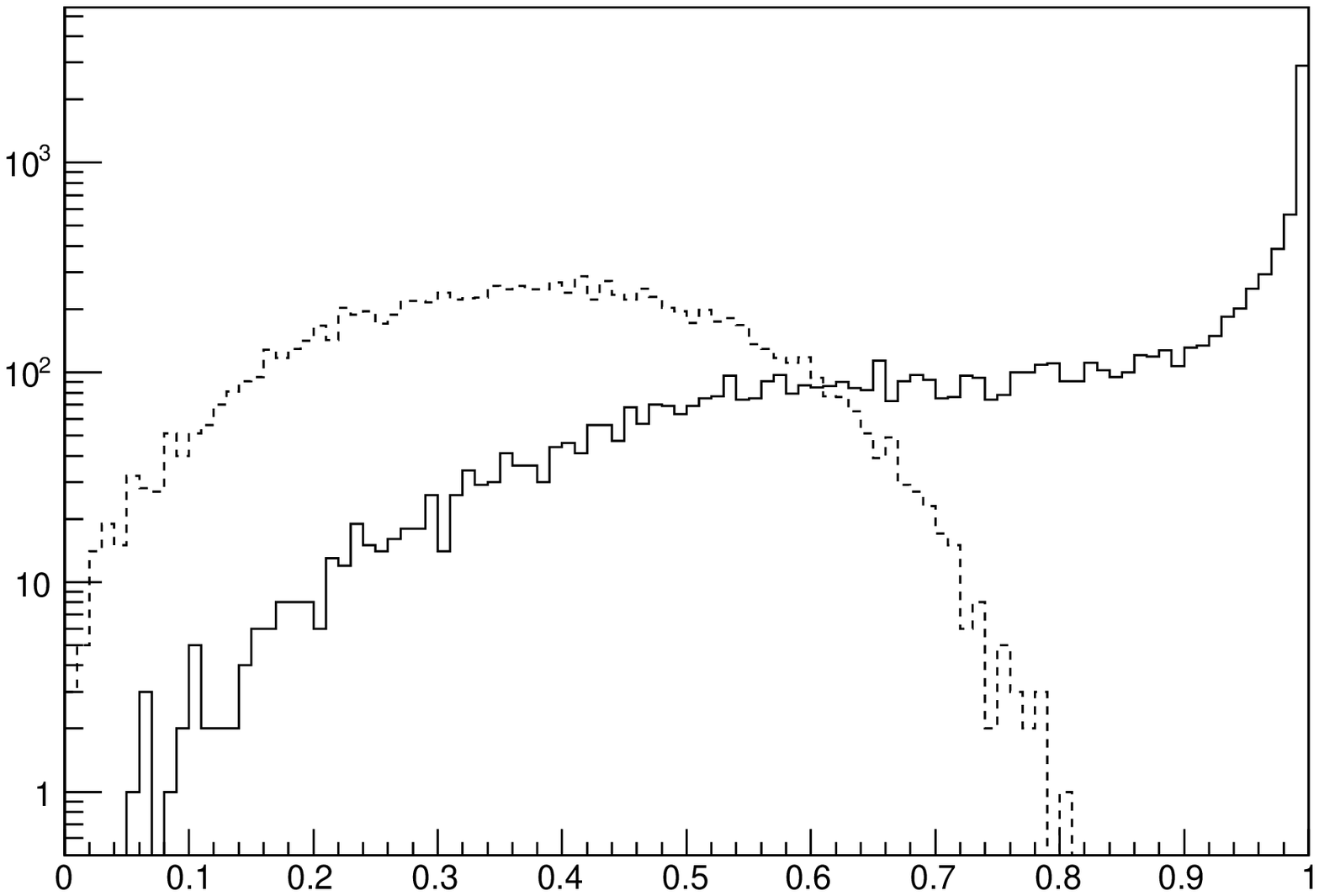,height=5cm,width=6.5cm}
&
\epsfig{figure=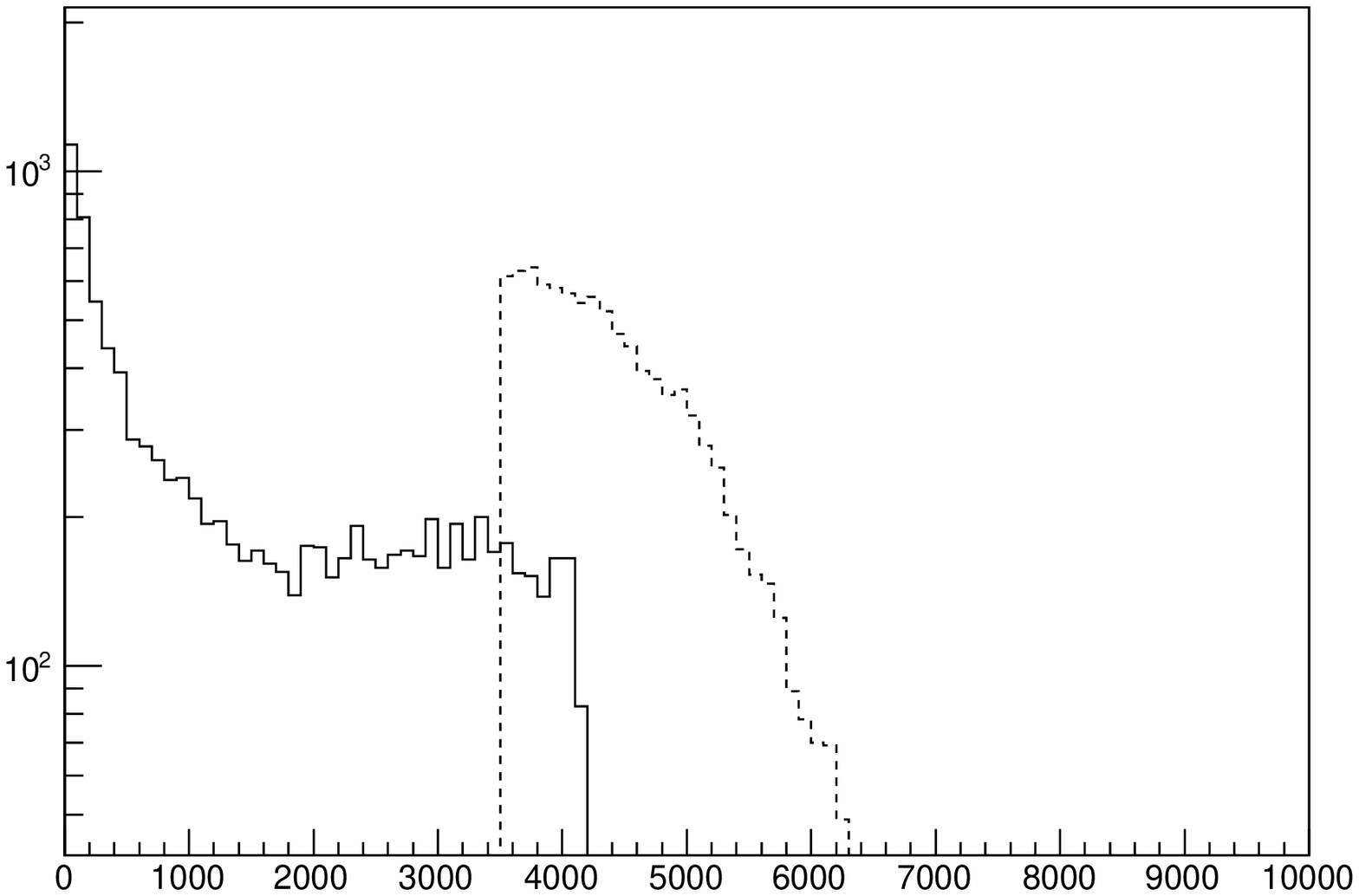,height=5cm,width=6.5cm}
\\
\small{$\beta_0$}
&
{\small{$M_0$}}
\end{tabular}
}
\caption{Distribution of speed $\beta_0$ (left panel) and mass $M_0$ (in GeV; right panel)
of the remnant black holes for KINCUT=TRUE (dashed line) and KINCUT=FALSE (solid line).
Both plots are for $\sqrt{s}=14\,$TeV with $M_{\rm G}=3.5\,$TeV and initial 
$M_{\rm BH}\ge 2\,M_{\rm G}$ in $D=6$ total dimensions and $10^4$ total black hole events.
\label{beta}}
\end{figure}
\begin{figure}[h!]
\centerline{\begin{tabular}{ cc}
\epsfig{figure=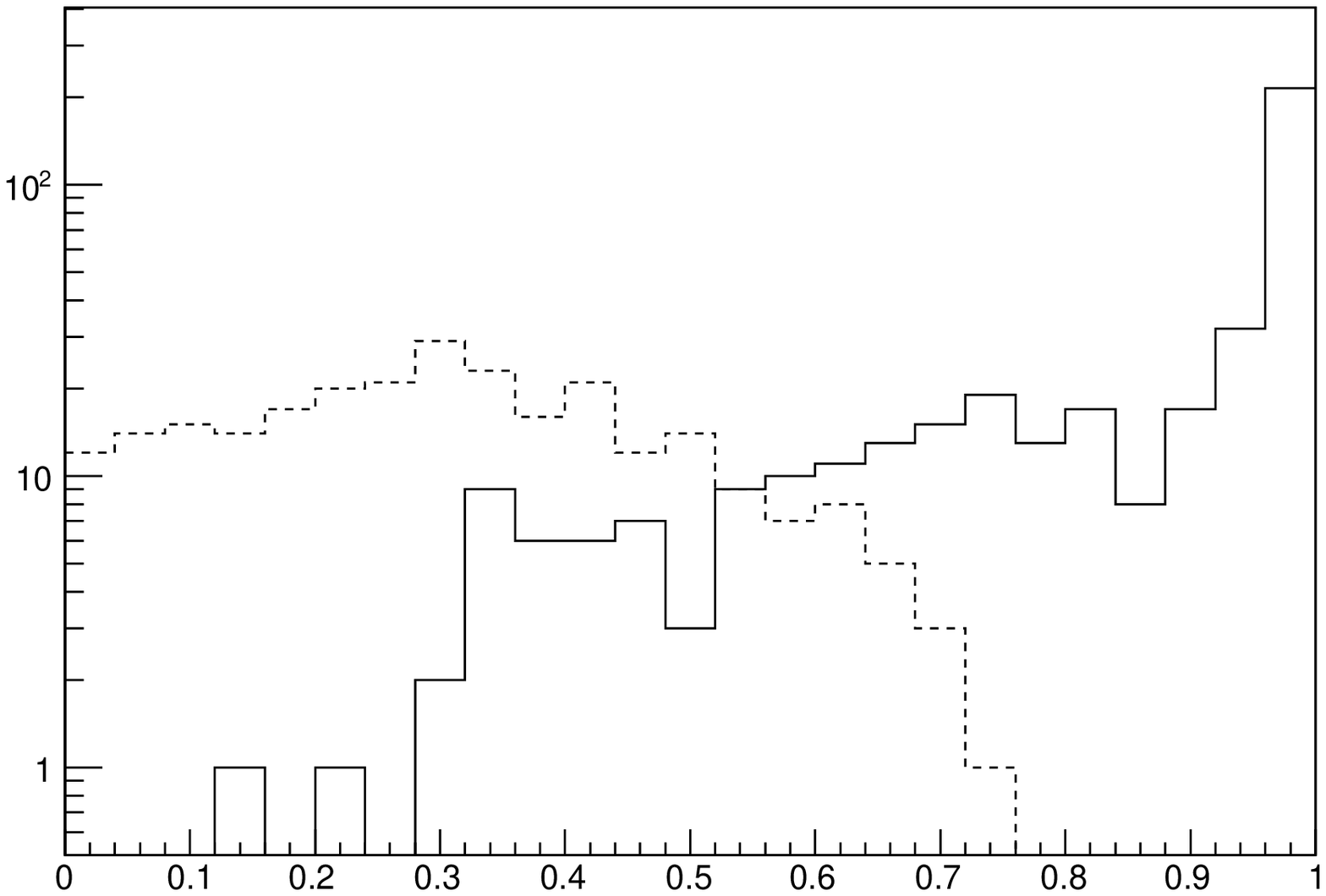,height=5cm,width=6.5cm}
&
\epsfig{figure=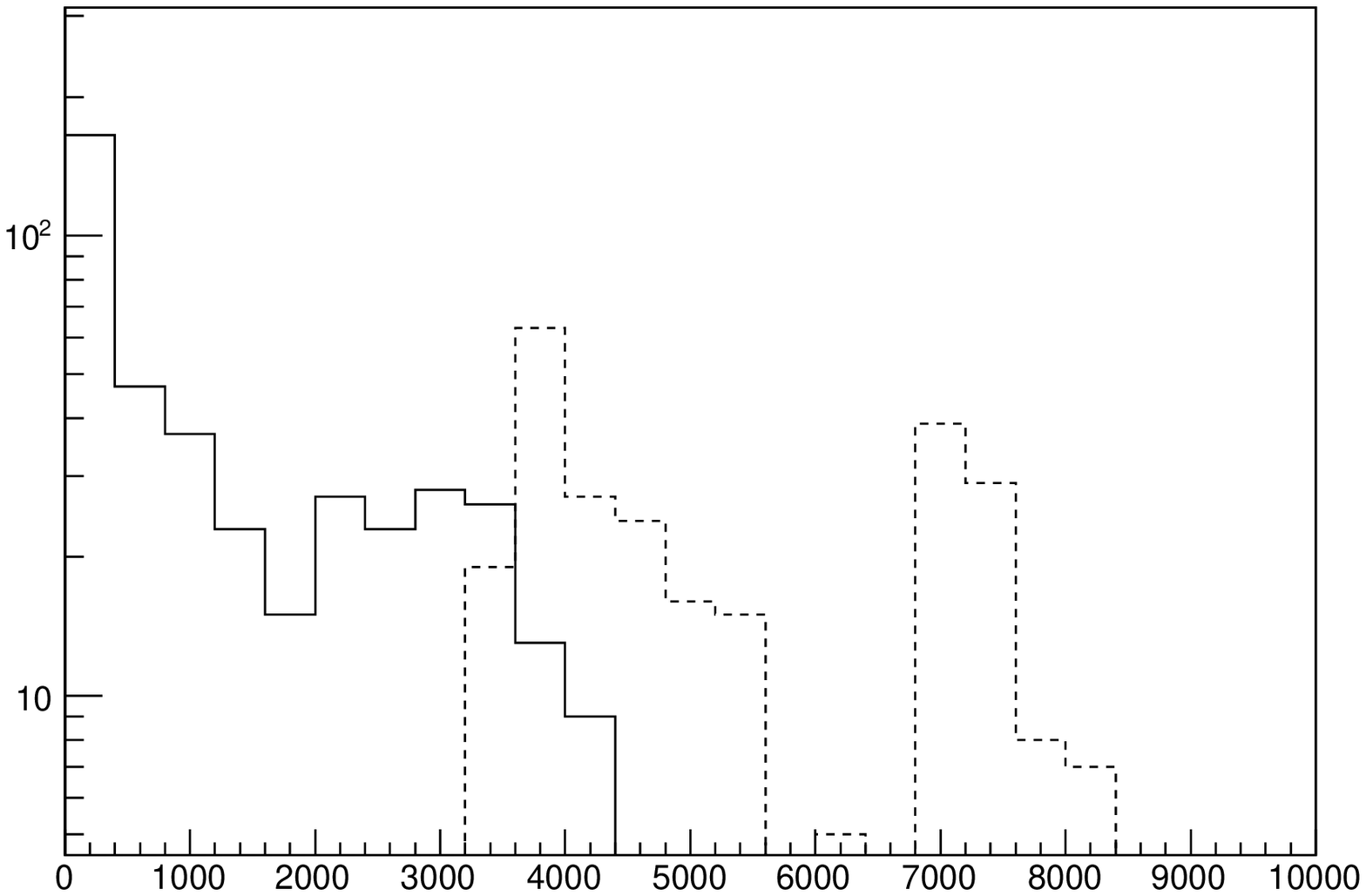,height=5cm,width=6.5cm}
\\
\small{$\beta_0$}
&
\small{$M_0$}
\end{tabular}}
\caption{Distribution of speed $\beta_0$ (left panel) and mass $M_0$ (in GeV; right panel)
of the charged remnant black holes for $M_0>M_{\rm G}$
(dashed line) and $M_0>0$ (solid line).
Both plots are for $\sqrt{s}=14\,$TeV with $M_{\rm G}=3.5\,$TeV in $D=6$
total dimensions and $10^4$ total events.
\label{betac}}
\end{figure}
\par
The numerical simulations show that the remnant black holes are expected
to have a typical speed $\beta_0=v_0/c$ with the distribution shown in the
left panel of Fig.~\ref{beta}, for a sample of $10^4$ black holes, where two
different scenarios for the end-point of the decay were assumed.
The dashed line represents the case when the decay
is prevented from producing a remnant with proper mass $M_0$ below $M_{\rm G}$
(but could stop at $M_0>M_{\rm G}$), whereas the solid line
represents black hole remnants produced when the last emission
is only required to keep $M_0>0$.
The mass $M_0$ for the remnants in the two cases is distributed
according to the plots in the right panel of Fig.~\ref{beta}.
In the former case, with the remnant mass $M_0\gtrsim M_{\rm G}$,
a smaller amount of energy is emitted before the hole
becomes a remnant, whereas in the latter much lighter remnants 
are allowed.
The first scenario provides a better description for black hole remnants
resulting from the partial decay of quantum black holes, and
the second scenario is mostly presented for the sake of completeness.
\par
The same quantities, speed $\beta_0$ and mass $M_0$, but only for
the charged remnants, are displayed in Fig.~\ref{betac}, again for
a sample of $10^4$ black hole events. 
The left panel shows that, including both scenarios,
one can expect the charged remnant velocity is quite evenly distributed
on the entire allowed range, but $\beta_0$ is generally smaller when
the remnant mass is larger than $M_{\rm G}$.
As it was shown earlier, black hole remnants are likely to have masses
of the order of $M_{\rm G}$ or larger, therefore from now on we will focus
on this case only. 
\begin{figure}[t]
\centerline{\begin{tabular}{cc}
\raisebox{3cm}{\small{$\beta_0$}}
\epsfig{figure=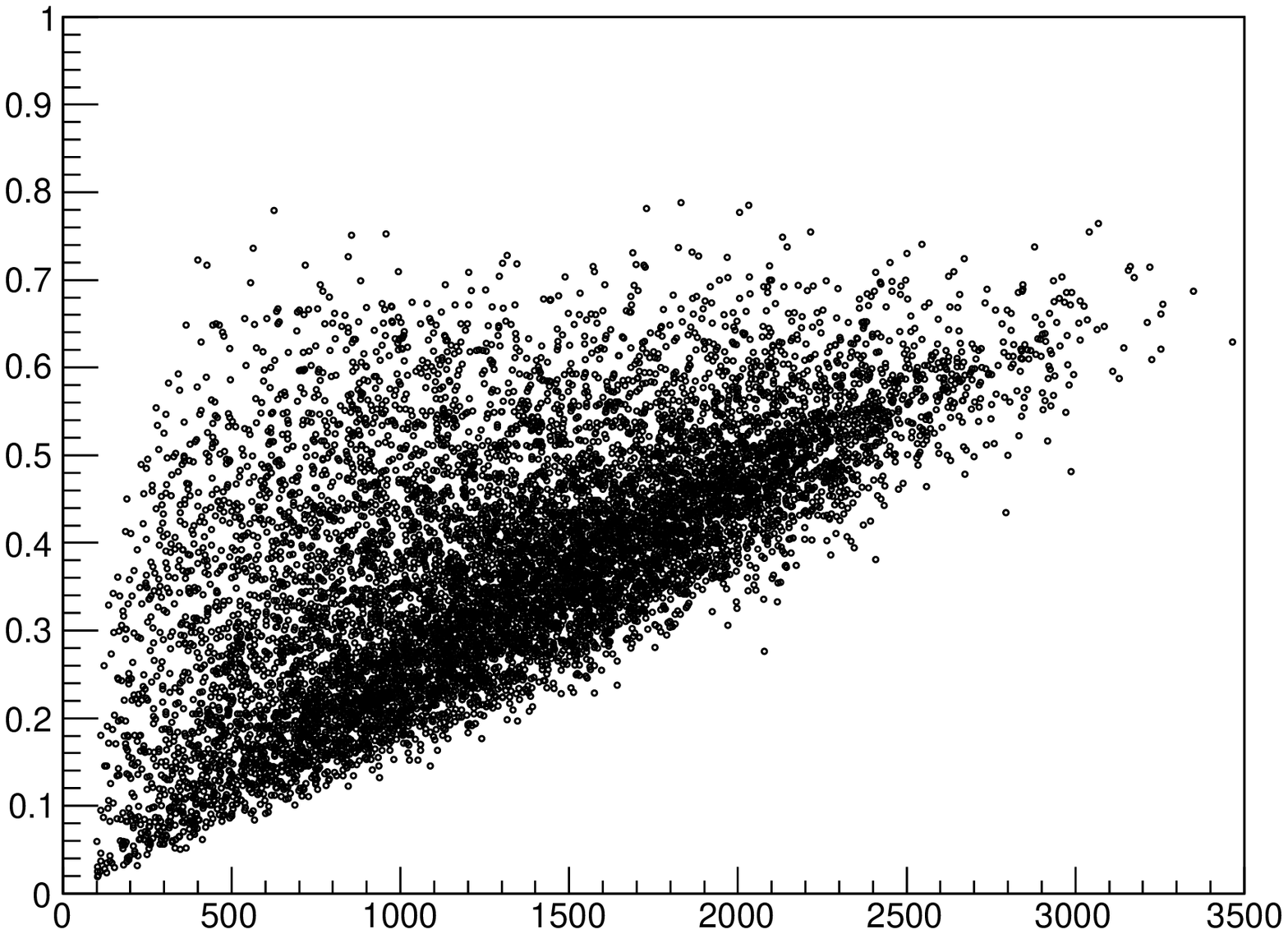,height=5cm,width=6.5cm}
&
\raisebox{3cm}{\small{$\beta_0$}}
\epsfig{figure=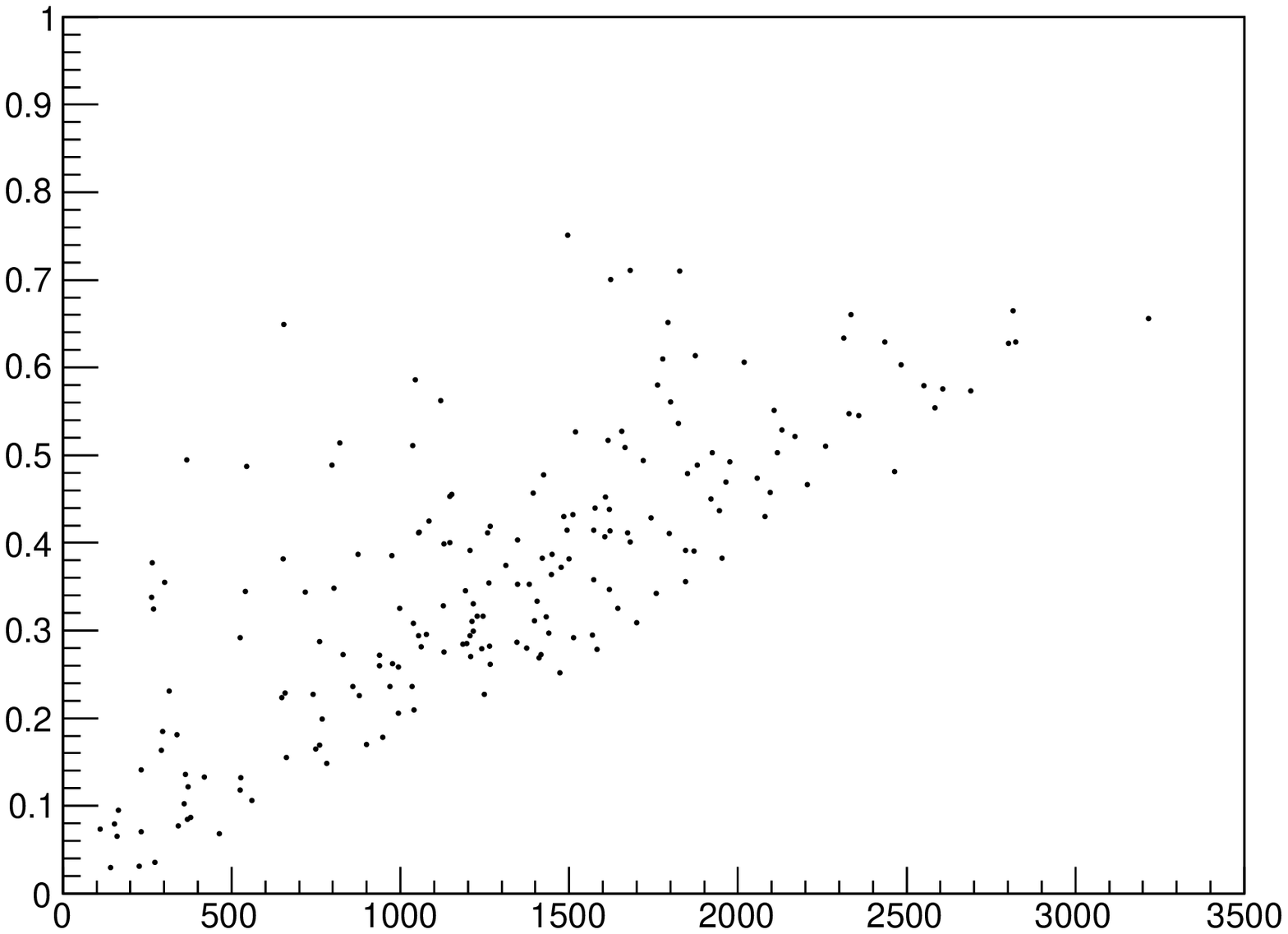,height=5cm,width=6.5cm}
\\
{\small{$P_T$}}
&
{\small{$P_T$}}
\end{tabular}}
\caption{Distribution of $\beta_0$ vs $P_T$ (in GeV) with $M_0>M_{\rm G}$
for neutral remnants (left panel) and charged remnants (right panel) for $P_T>100\,$GeV.
Both plots are for $\sqrt{s}=14\,$TeV with $M_{\rm G}=3.5\,$TeV in $D=6$ total dimensions
and $10^4$ total events.
\label{scatterBetaT}}
\end{figure}
\par
\begin{figure}[t!]
\centerline{\begin{tabular}{cc}
\raisebox{3cm}{\small{$\beta$}}
\epsfig{figure=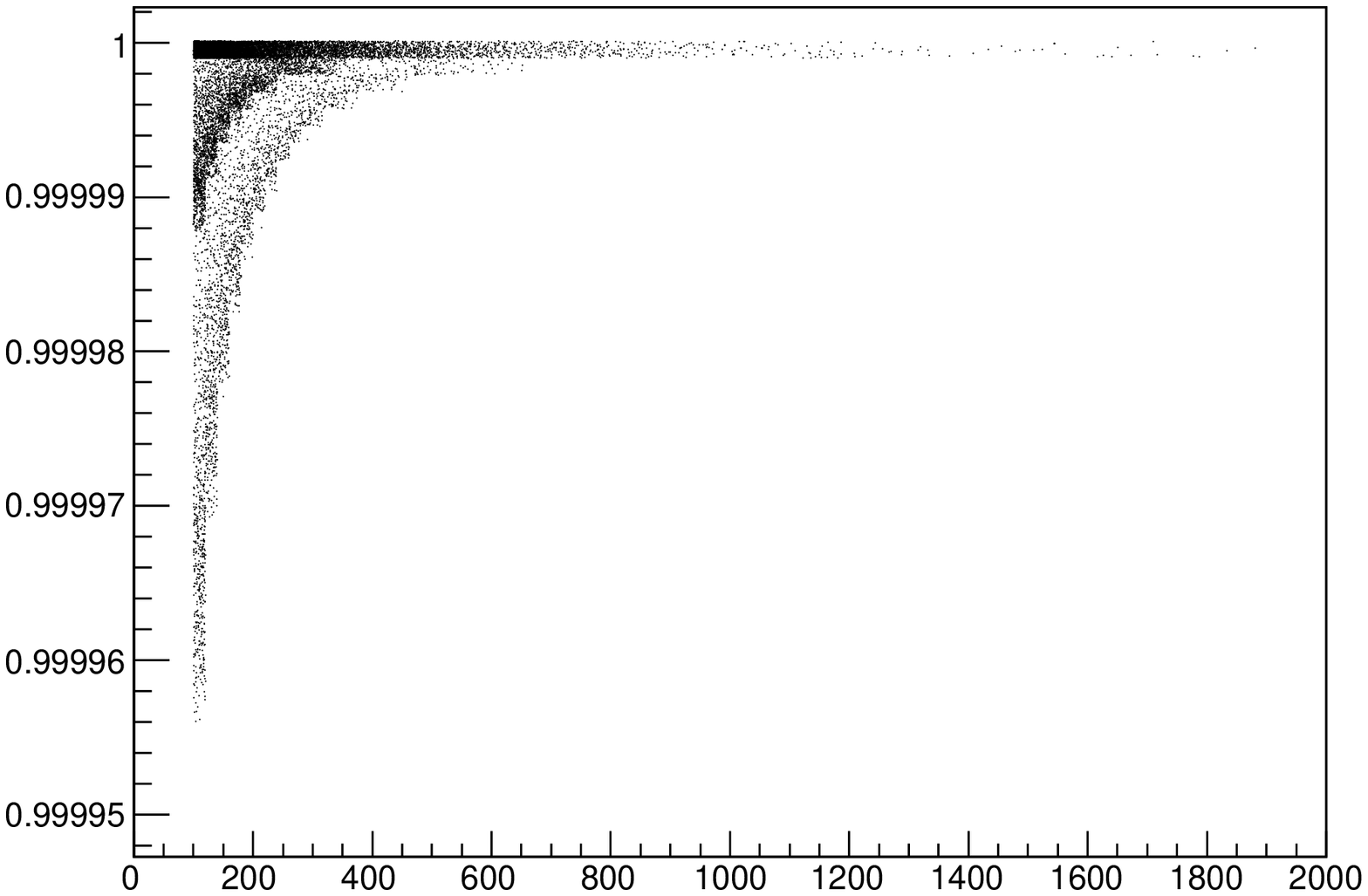,height=5cm,clip=}
&
\raisebox{3cm}{\small{$\beta$}}
\epsfig{figure=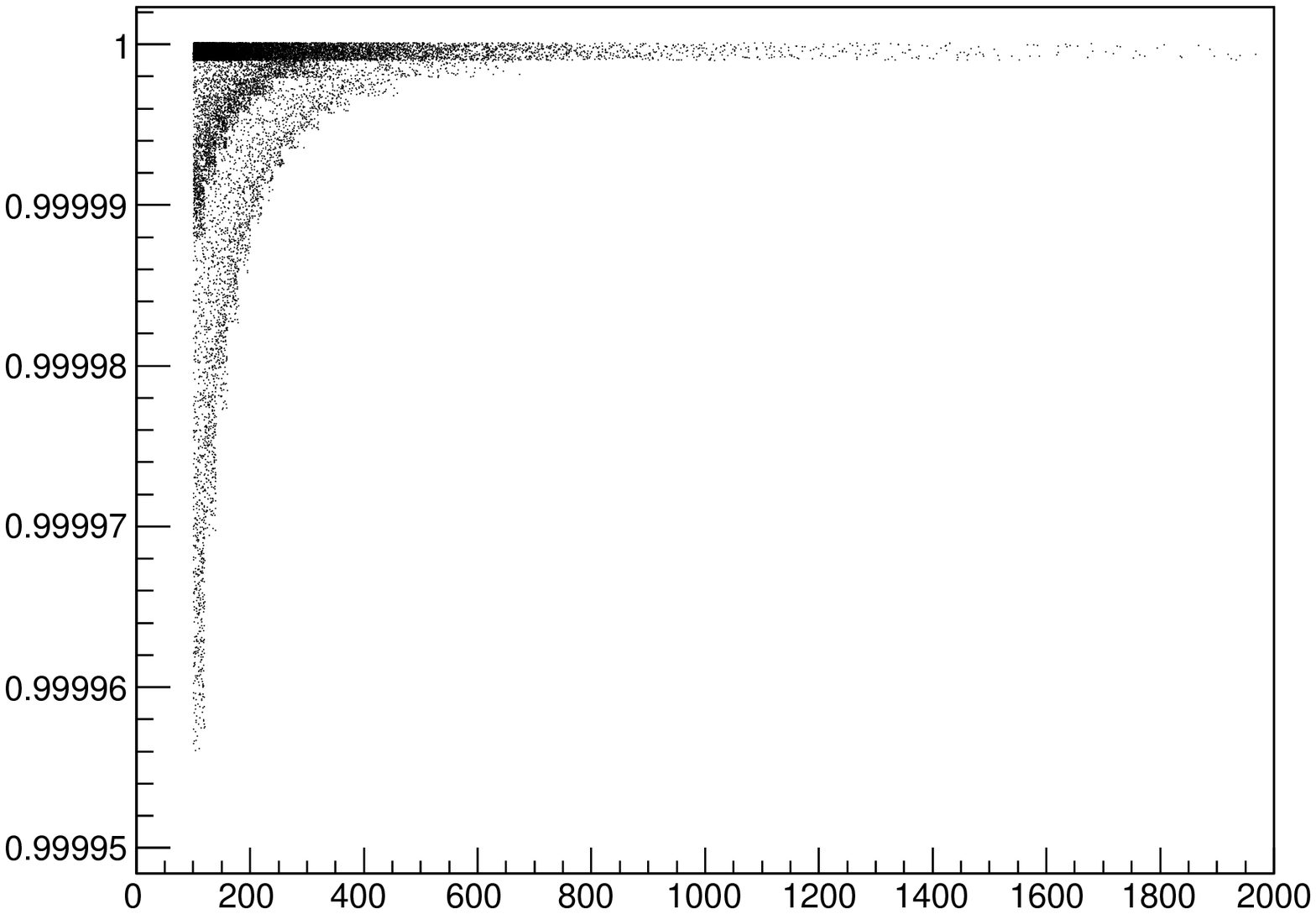,height=5cm,clip=}
\\
{\small{$P_T$}}
&
{\small{$P_T$}}
\end{tabular}}
\caption{Distribution of $\beta$ vs $P_T$ (in GeV) for background particles with $P_T>100\,$GeV,
in events with remnant black holes and $M_0>M_{\rm G}$ (left panel) or $M_0>0$ (right panel).
Both plots are for $\sqrt{s}=14\,$TeV with $M_{\rm G}=3.5\,$TeV in $D=6$
total dimensions and $10^4$ total events.
\label{scatterBetaNo}}
\end{figure}
\begin{figure}[h!]
\centerline{\begin{tabular}{c}
\raisebox{3cm}{\small{$\beta$}}
\epsfig{figure=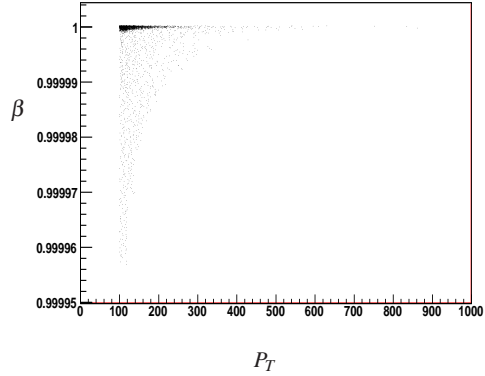,height=5cm,width=6.5cm}
\\
{\small{$P_T$}}
\end{tabular}}
\caption{Distribution of $\beta$ vs $P_T$ (in GeV) for particles with $P_T>100\,$GeV,
in events with $t\,\bar t$ for $\sqrt{s}=14\,$TeV.
\label{scatterBetatt}}
\end{figure}

For phenomenological reasons, it is very instructive to consider the distribution of the speed
$\beta_0$ with respect to transverse momenta $P_T$ for remnant black holes.
A cut-off is set for particles with transverse momentum of $P_T>100\,$GeV. 
Fig.~\ref{scatterBetaT} shows separately the distributions of $\beta_0$ for neutral
and charged remnants.
We first recall that the remnant velocities are lower because the masses of remnant
black holes in this case are typically larger. 
Fig.~\ref{scatterBetaNo} shows the similar plot $\beta_0$ versus $P_T$ for the
background particles. 
When comparing the two plots, remnants appear clearly distinguished 
since there is hardly any black hole with $\beta_0\gtrsim 0.7$, whereas all the
background particles have $\beta\simeq 1$.
The speeds $\beta_0$ of the remnants can also be compared with the distributions
of $\beta$ for the $t\,\bar t$ process (which can be considered as one of the main
backgrounds) shown in Fig.~\ref{scatterBetatt}.
Taking into account the production cross section $\sigma_{t\,\bar t}(14\,{\rm TeV})\simeq 880\,$pb,
and the branching ratio for single-lepton decays
(final states with significant missing transverse energy), for a luminosity of $L=10\,$fb$^{-1}$
a number of $3.9\times 10^6$ such events are expected.
This must be compared with the expected number of $400$ black hole events
that could be produced for the same luminosity.
\par
\begin{figure}[t!]
\centerline{\begin{tabular}{cc}
\raisebox{3cm}{\small{$\frac{dE}{dx}$}}
\epsfig{figure=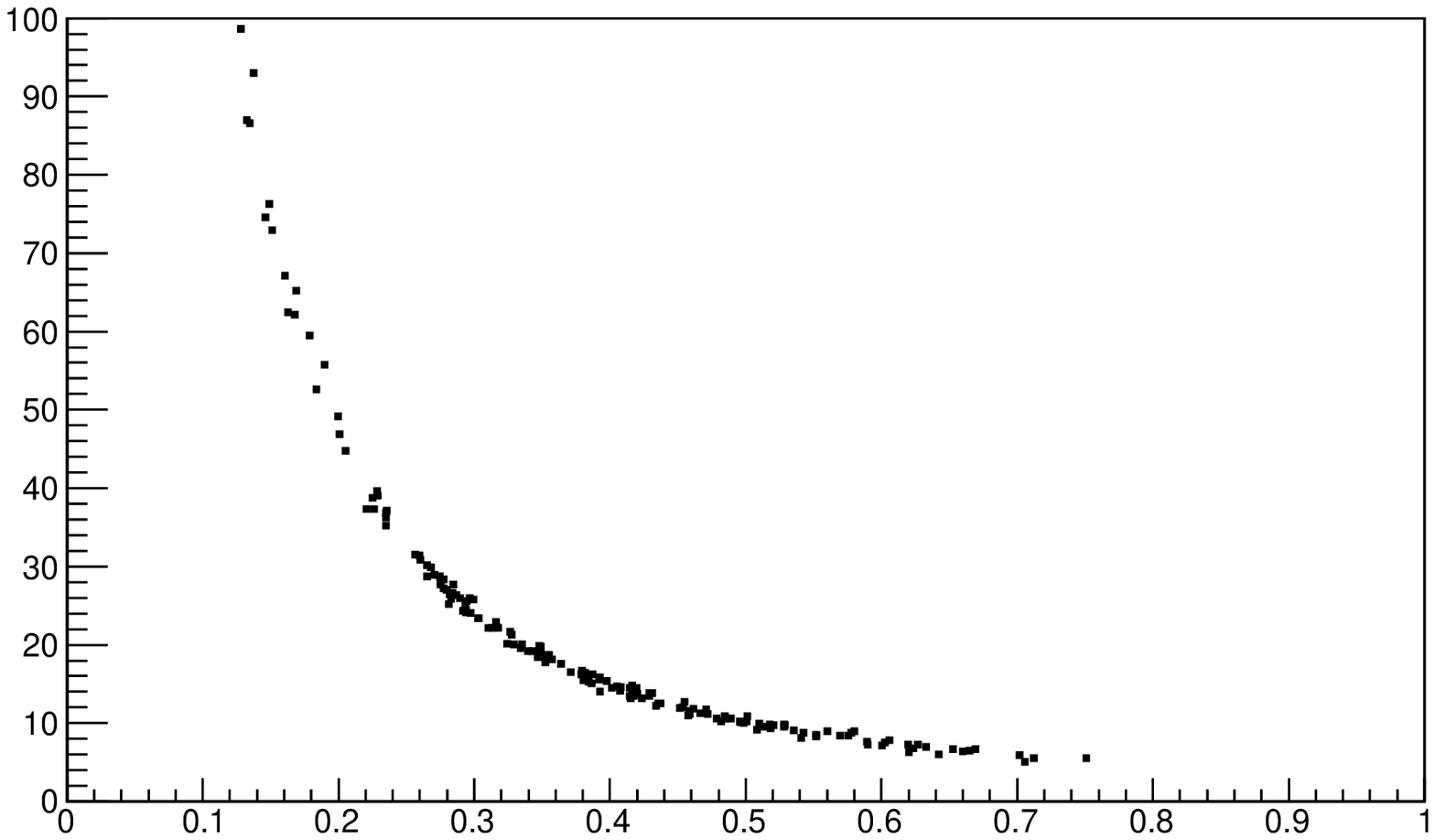,height=5cm,width=6.5cm}
&
\raisebox{3cm}{\small{$\frac{dE}{dx}$}}
\epsfig{figure=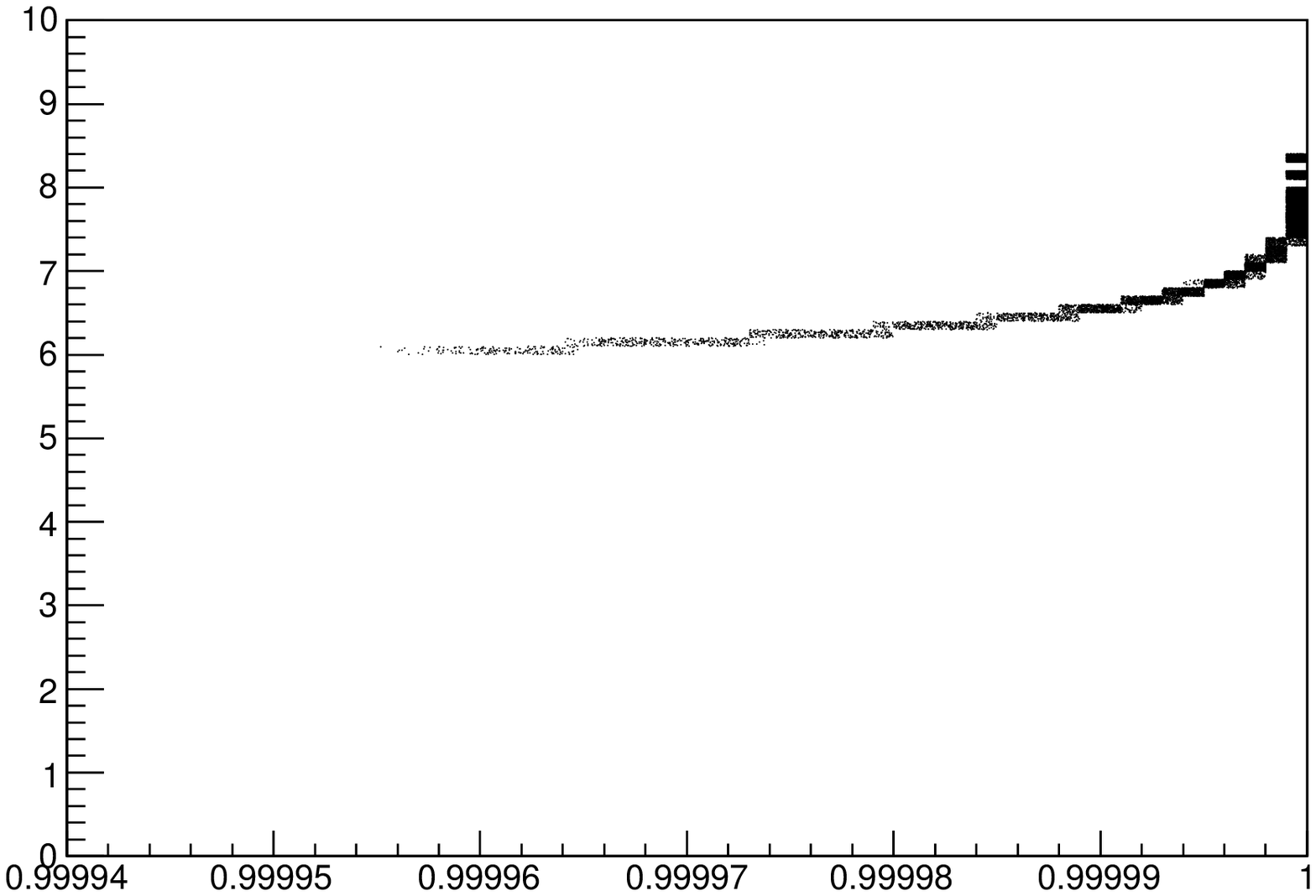,height=5cm,width=6.5cm}
\\
{\small{$\beta_0$}}
&
{\small{$\beta$}}
\end{tabular}}
\caption{Typical energy loss per unit distance (in MeV/cm) from charged remnant black holes 
vs $\beta_0$, for $M_0>M_{\rm G}$ (left panel) and analogous quantity for background particles
(right panel).
Both plots are for $\sqrt{s}=14\,$TeV with $M_{\rm G}=3.5\,$TeV in $D=6$
total dimensions and $10^4$ total events.
}
\label{dEdxT}
\end{figure}
\begin{figure}[h!]
\centerline{\begin{tabular}{c}
\raisebox{3cm}{\small{$\frac{dE}{dx}$}}
\epsfig{figure=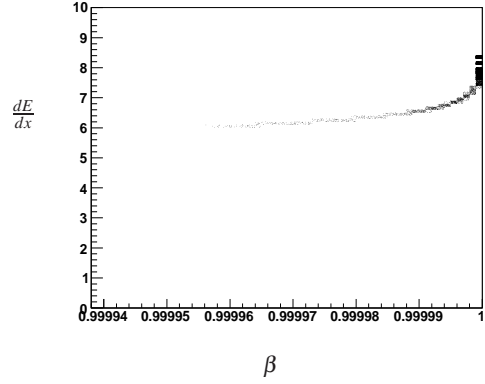,height=5cm,width=6.5cm}
\\
{\small{$\beta$}}
\end{tabular}}
\caption{Typical energy loss per unit distance (in MeV/cm) from charged particles vs $\beta$
in $10^4$ total events with $t\,\bar t$ at $\sqrt{s}=14\,$TeV.
}
\label{dEdxtt}
\end{figure}

Charged particles also release energy when traveling through a medium.
The energy released by a particle of mass $M$ and charge $Q=z\,e$
can be estimated using the well-known Bethe-Bloch equation.
For particles moving at relativistic speeds, one has an energy loss per
distance travelled given by
\be
\frac{dE}{dx}
=
-4\,\pi\,N_A\,r_e^2\,m_e\,c^2\,
\frac{Z\,\rho}{A\,\beta^2}
\left[
\ln\left(\frac{2\,m_e\,c^2\,\beta^2}{I}\right)
-\beta^2-\frac{\delta}{2}\right]
\ ,
\label{bethe}
\ee
where $N_A$ is Avogadro's number, 
$m_e$ and $r_e$ the electron mass and classical radius,
$Z$, $A$ and $\rho$ the atomic number, atomic weight and density of the medium,
$I \simeq 16\,Z^{0.9}\,{\rm eV}$ its mean excitation potential,
and $\delta$ a constant that describes the screening of the electric field
due to medium polarisation.
For the LHC, one can use the values for Si, as the $dE/dX$ can be effectively
measured in the ATLAS Inner Detector, namely 
$\rho=2.33\,$g$/$cm$^3$, $Z= 14$, $A = 28$, $I = 172\,$eV and
$\delta=0.19$.
On using the $\beta_0$ for charged remnant black holes from the right panel of
Fig.~\ref{scatterBetaT}, one then obtains the typical
distributions displayed in Figs.~\ref{dEdxT}, where
the energy loss from remnant black holes is compared with analogous quantities
for ordinary particles coming from black hole evaporation.
One can then also compare with the energy loss in $t\,\bar t$ events
displayed in Fig.~\ref{dEdxtt}.
It can be seen that a cut around $10\,$MeV$/$cm would clearly isolate
remnants black holes, since they would mostly loose more energy.
\par
The charged stable remnants behave as massive muons, travelling long distances through
the detector and releasing only a negligible fraction of their total energy.
The main problem in detecting such states at the LHC is the trigger time width of $25\,$ns
(1 bunch crossing time). 
Due to their low speed, most of them will reach the muon system out of time and could not
be accepted by the trigger.
A study performed at ATLAS set a threshold cut of $\beta > 0.62$ in order to have a muon
trigger in the event (slower particles end up out of the trigger time window).
In order to access the low $\beta$ range, one can imagine to trigger on the missing transverse
energy ($E_T^{\rm miss}$), copiously produced by the charged remnants, or on other standard
particles produced in the black hole evaporation (typically electrons or muons).
Another possibility is to trigger on ordinary particles, typically electrons or muons with
high transverse momentum $P_T$, in order to reduce the high potential background
coming from QCD multi-jet events.
Once the events have been accepted by the trigger the signal has to be isolated from
the background by means of the $dE/dX$ measurement.
\section{Concluding remarks}
We have seen that the very existence of black holes in gravity is at the heart of GUPs for
quantum mechanics, which imply the existence of a minimum measurable length.
These modifications of quantum mechanics, in turn, imply that black holes can only exist
above a minimum mass threshold.
Minimum mass black holes could be stable, or metastable remnants with zero Hawking
temperature.
In any case, they would belong to the realm of quantum objects, for which we still have
limited theoretical understanding.
\par
In four-dimensional gravity, this minimum mass is usually predicted to be of the order of
the Planck mass, $\mpl\simeq 10^{16}\,$TeV, well above the energies that can be
reached in our laboratories.
However, if the universe really contains extra spatial dimensions hidden to our
direct investigation, the fundamental gravitational mass could be much lower
and potentially within the reach of our experiments.
Black holes might therefore be produced in future colliders, and deviations from
the standard uncertainty relations of quantum mechanics might be testable at length
scales much larger than the Planck length, $\lp\simeq 10^{-35}\,$m.
\par
All of the above considered, black holes and a minimum measurable length scale are at the 
very frontiers of contemporary fundamental physics.
%
%
%
%

%
%

\begin{thebibliography}{99}
%
\bibitem{Adl10} 
  R.~J.~Adler,
  \textit{Am.\ J.\ Phys.\  }{\bf 78}, 925 (2010).
  
\bibitem{KMM95} 
  A.~Kempf, G.~Mangano and R.~B.~Mann,
  \textit{Phys.\ Rev.\ D} {\bf 52}, 1108 (1995).
  
\bibitem{ACS01} 
  R.~J.~Adler, P.~Chen and D.~I.~Santiago,
  \textit{Gen.\ Rel.\ Grav.\  }{\bf 33}, 2101 (2001).
  
\bibitem{ACV08} 
  D.~Amati, M.~Ciafaloni and G.~Veneziano,
\textit{  JHEP }{\bf 02}, 049 (2008).


\bibitem{DGG11} 
  G.~Dvali, G.~F.~Giudice, C.~Gomez and A.~Kehagias,
  \textit{JHEP} {\bf 08}, 108 (2011).
 
\bibitem{AuS13b} 
  A.~Aurilia and E.~Spallucci,
  ``Planck's uncertainty principle and the saturation of Lorentz boosts by Planckian black holes,''
  arXiv:1309.7186 [gr-qc].
  
  
\bibitem{Hos12} 
  S.~Hossenfelder,
  \textit{Living Rev.\ Rel.\ } {\bf 16}, 2 (2013).
  
\bibitem{DeW62} 
  B.~S.~DeWitt, ``The quantization of geometry,''
 in Louis Witten (ed.), Gravitation: An Introduction to Current Research, New York: J. Wiley and Sons, pp. 266-381 (1962).
  
\bibitem{ACV87} 
  D.~Amati, M.~Ciafaloni and G.~Veneziano,
\textit{  Phys.\ Lett.\ B} {\bf 197}, 81 (1987).
  

\bibitem{Ven86} 
  G.~Veneziano,
  \textit{Europhys.\ Lett.\ } {\bf 2}, 199 (1986).

\bibitem{ACV89} 
  D.~Amati, M.~Ciafaloni and G.~Veneziano,
  \textit{Phys.\ Lett.\ B} {\bf 216}, 41 (1989).
  
 
\bibitem{Yon89} 
  T.~Yoneya,
 \textit{ Mod.\ Phys.\ Lett.\ A }{\bf 4}, 1587 (1989).
  
\bibitem{KPP90} 
  K.~Konishi, G.~Paffuti and P.~Provero,
  \textit{Phys.\ Lett.\ B} {\bf 234}, 276 (1990).

\bibitem{AuS13} 
  A.~Aurilia and E.~Spallucci,
 \textit{ Adv.\ High Energy Phys.\ } {\bf 2013}, 531696 (2013).

\bibitem{RoS95} 
  C.~Rovelli and L.~Smolin,
 \textit{ Nucl.\ Phys.\ B }{\bf 442}, 593 (1995)
  [\textit{Erratum-ibid.\ B }{\bf 456}, 753 (1995)].

\bibitem{SeW99} 
  N.~Seiberg and E.~Witten,
 \textit{ JHEP} {\bf 09}, 032 (1999).
  
\bibitem{Wei80} 
  S.~Weinberg,
  ``Ultraviolet divergences in quantum theories of gravitation,'' in General Relativity: an Einstein centenary survey, ed. S. W. Hawking and W. Israel. Cambridge University Press. pp. 790-831.
  
\bibitem{Reu96} 
  M.~Reuter,
  \textit{Phys.\ Rev.\ D} {\bf 57}, 971 (1998).


\bibitem{Ame94} 
  G.~Amelino-Camelia,
  \textit{Mod.\ Phys.\ Lett.\ A }{\bf 9}, 3415 (1994).

  
\bibitem{Gar95} 
  L.~J.~Garay,
  \textit{Int.\ J.\ Mod.\ Phys.\ A} {\bf 10}, 145 (1995).


\bibitem{SNB12} 
  M.~Sprenger, P.~Nicolini and M.~Bleicher,
  \textit{Eur.\ J.\ Phys.\ } {\bf 33}, 853 (2012).

\bibitem{SCgup}
F.~Scardigli and R.~Casadio,
\textit{ Int.\ J.\ Mod.\ Phys.\ D} {\bf 18} (2009) 319.

%
\bibitem{hoop} 
K.S.~Thorne,
\textit{Nonspherical gravitational collapse: A short review},
in J.R.~Klauder, \textit{Magic Without Magic}, San Francisco (1972), 231.
%
\bibitem{Casadio:2013tma} 
  R.~Casadio,
 ``Localised particles and fuzzy horizons: A tool for probing Quantum Black Holes,''
  arXiv:1305.3195 [gr-qc].
  %
\bibitem{Casadio:2013tza} 
  R.~Casadio,
 ``What is the Schwarzschild radius of a quantum mechanical particle?,''
  arXiv:1310.5452 [gr-qc].
  %
\bibitem{Casadio:2013aua} 
  R.~Casadio and F.~Scardigli,
  \textit{Eur.\ Phys.\ J.\ C} {\bf 74}, 2685 (2014).
%
\bibitem{Qhoop}
R.~Casadio, O.~Micu and F.~Scardigli,
\textit{Phys. Lett.} {\bf B 732} (2014) 105.
%
\bibitem{Cminmass}
R.~Casadio and J.~Ovalle,
\textit{Gen.\ Rel.\ Grav.} {\bf 46} (2014) 1669.
%
\bibitem{Cminmass1}
R.~Casadio and J.~Ovalle,
\textit{Phys.\ Lett.\ B} {\bf 715} (2012) 251.
%
\bibitem{scale}
G.~L.~Alberghi, R.~Casadio, O.~Micu, A.~Orlandi,
\textit{JHEP} {\bf  09 }, 023 (2011).
\bibitem{BeR07} 
  M.~Berkooz and D.~Reichmann,
 \textit{ Nucl.\ Phys.\ Proc.\ Suppl.\ } {\bf 171}, 69 (2007).

\bibitem{Rov08} 
  C.~Rovelli,
  \textit{Living Rev.\ Rel.\ } {\bf 1}, 1 (1998); 
  \textit{Living Rev.\ Rel.\ } {\bf 11}, 5 (2008), Chapt.~8, pp~42.
  %
\bibitem{bardeen}
J. M. Bardeen, ``Non-singular general-relativistic gravitational collapse," 
in Proceedings of International Conference GR5, p. 174, USSR, Tbilisi, Georgia, 1968.
%
\bibitem{aa1} 
  A.~Aurilia, G.~Denardo, F.~Legovini and E.~Spallucci,
 \textit{ Phys.\ Lett.\ B }{\bf 147}, 258 (1984).
  %
  \bibitem{aa2} 
  A.~Aurilia, G.~Denardo, F.~Legovini and E.~Spallucci,
 \textit{ Nucl.\ Phys.\ B }{\bf 252}, 523 (1985).

\bibitem{aa3} 
  A.~Aurilia, R.~S.~Kissack, R.~B.~Mann and E.~Spallucci,
\textit{  Phys.\ Rev.\ D }{\bf 35}, 2961 (1987).
  
  \bibitem{fmm} 
  V.~P.~Frolov, M.~A.~Markov and V.~F.~Mukhanov,
\textit{  Phys.\ Rev.\ D }{\bf 41}, 383 (1990).
  
\bibitem{beato} 
  E.~Ayon-Beato and A.~Garcia,
 \textit{ Phys.\ Rev.\ Lett.\ } {\bf 80}, 5056 (1998).

\bibitem{dym}
 I.G.~Dymnikova,
 \textit{Int. J. Mod. Phys. } {\bf D5}, 529 (1996)
%

\bibitem{dym2}
I.G.~Dymnikova,
\textit{Int. J. Mod. Phys. } {\bf D12}, 1015 (2003)

\bibitem{MbK05} 
  M.~R.~Mbonye and D.~Kazanas,
 \textit{ Phys.\ Rev.\ D }{\bf 72}, 024016 (2005).

\bibitem{MbK08} 
  M.~R.~Mbonye and D.~Kazanas,
  \textit{Int.\ J.\ Mod.\ Phys.\ D} {\bf 17}, 165 (2008).


  \bibitem{hay} 
  S.~A.~Hayward,
 \textit{ Phys.\ Rev.\ Lett.\ } {\bf 96}, 031103 (2006).
  
\bibitem{SpS12} 
  E.~Spallucci and A.~Smailagic,
  \textit{Phys.\ Lett.\ B }{\bf 709}, 266 (2012).

\bibitem{NiS14} 
  P.~Nicolini and E.~Spallucci,
  \textit{Adv.\ High Energy Phys.\ } {\bf 2014}, 805684 (2014).


\bibitem{Mod06} 
  L.~Modesto,
 \textit{ Class.\ Quant.\ Grav.\ } {\bf 23}, 5587 (2006).
  
\bibitem{MoP09} 
  L.~Modesto and I.~Premont-Schwarz,
 \textit{ Phys.\ Rev.\ D} {\bf 80}, 064041 (2009).
  
\bibitem{BoR00} 
  A.~Bonanno and M.~Reuter,
  \textit{Phys.\ Rev.\ D} {\bf 62}, 043008 (2000).
  
\bibitem{Car13} 
  B.~J.~Carr,
  \textit{Mod.\ Phys.\ Lett.\ A} {\bf 28}, 1340011 (2013).

\bibitem{CMP11} 
  B.~Carr, L.~Modesto and I.~Premont-Schwarz,
  ``Generalized Uncertainty Principle and Self-dual Black Holes,''
  arXiv:1107.0708 [gr-qc].

  
\bibitem{stefano} 
  S.~Ansoldi,
  ``Spherical black holes with regular center: a review of existing models including a recent realization with Gaussian sources,'' in Proceedings of Conference on Black Holes and Naked Singularities, 10-12 May 2007, Milan, Italy, arXiv:0802.0330 [gr-qc].

\bibitem{Nic09} 
  P.~Nicolini,
  \textit{Int.\ J.\ Mod.\ Phys.\ A} {\bf 24}, 1229 (2009).

\bibitem{BaN93} 
  H.~Balasin and H.~Nachbagauer,
 \textit{ Class.\ Quant.\ Grav.\ } {\bf 10}, 2271 (1993).

\bibitem{BaN94} 
  H.~Balasin and H.~Nachbagauer,
 \textit{ Class.\ Quant.\ Grav.\ } {\bf 11}, 1453 (1994).



\bibitem{SmS03a} 
  A.~Smailagic and E.~Spallucci,
  \textit{J.\ Phys.\ A} {\bf 36}, L467 (2003).

\bibitem{SmS03b} 
  A.~Smailagic and E.~Spallucci,
  \textit{J.\ Phys.\ A} {\bf 36}, L517 (2003).

\bibitem{SmS04} 
  A.~Smailagic and E.~Spallucci,
  \textit{J.\ Phys.\ A} {\bf 37}, 1 (2004)
  [\textit{Erratum-ibid.\ A} {\bf 37}, 7169 (2004)]

\bibitem{SSN06} 
  E.~Spallucci, A.~Smailagic and P.~Nicolini,
  \textit{Phys.\ Rev.\ D} {\bf 73}, 084004 (2006).
  
\bibitem{KoN10} 
  M.~Kober and P.~Nicolini,
  \textit{Class.\ Quant.\ Grav.\ } {\bf 27}, 245024 (2010).
  
\bibitem{Casadio:2005vg} 
  R.~Casadio, P.~H.~Cox, B.~Harms and O.~Micu,
 \textit{ Phys.\ Rev.\ D} {\bf 73}, 044019 (2006).
  
\bibitem{Casadio:2007ec} 
  R.~Casadio, A.~Gruppuso, B.~Harms and O.~Micu,
 \textit{ Phys.\ Rev.\ D} {\bf 76}, 025016 (2007).
  
 
\bibitem{NSS05} 
  P.~Nicolini, A.~Smailagic and E.~Spallucci,
 \textit{ ESA Spec.\ Publ.\ } {\bf 637}, 11.1 (2006).


\bibitem{Nic05} 
  P.~Nicolini,
  \textit{J.\ Phys.\ A} {\bf 38}, L631 (2005)


\bibitem{NSS06} 
  P.~Nicolini, A.~Smailagic and E.~Spallucci,
  \textit{Phys.\ Lett.\ B} {\bf 632}, 547 (2006)

\bibitem{BGM10} 
  R.~Banerjee, S.~Gangopadhyay and S.~K.~Modak,
 \textit{ Phys.\ Lett.\ B }{\bf 686}, 181 (2010).


\bibitem{MMN11} 
  L.~Modesto, J.~W.~Moffat and P.~Nicolini,
  \textit{Phys.\ Lett.\ B} {\bf 695}, 397 (2011).
  
\bibitem{Mof11} 
  J.~W.~Moffat,
  \textit{Eur.\ Phys.\ J.\ Plus} {\bf 126}, 43 (2011).

\bibitem{Nic12} 
  P.~Nicolini,
  ``Nonlocal and generalized uncertainty principle black holes,''
  arXiv:1202.2102 [hep-th].

\bibitem{IMN13} 
  M.~Isi, J.~Mureika and P.~Nicolini,
  \textit{JHEP }{\bf 1311}, 139 (2013).
  
\bibitem{Mod12} 
  L.~Modesto,
  \textit{Phys.\ Rev.\ D} {\bf 86}, 044005 (2012).
  
\bibitem{BGK12} 
  T.~Biswas, E.~Gerwick, T.~Koivisto and A.~Mazumdar,
  \textit{Phys.\ Rev.\ Lett.\ } {\bf 108}, 031101 (2012).


\bibitem{CaO13} 
  R.~Casadio and A.~Orlandi,
  \textit{JHEP} {\bf 08}, 025 (2013).

\bibitem{DvG11} 
  G.~Dvali and C.~Gomez,
 \textit{ Fortsch.\ Phys.\  }{\bf 59}, 579 (2011).

\bibitem{DvG13a} 
  G.~Dvali and C.~Gomez,
\textit{  Fortsch.\ Phys.\  }{\bf 61}, 742 (2013).

\bibitem{DvG13b} 
  G.~Dvali and C.~Gomez,
 \textit{ Phys.\ Lett.\ B} {\bf 719}, 419 (2013).

\bibitem{DFG13} 
  G.~Dvali, D.~Flassig, C.~Gomez, A.~Pritzel and N.~Wintergerst,
  \textit{Phys.\ Rev.\ D} {\bf 88}, 124041 (2013).

\bibitem{BaN10} 
  D.~Batic and P.~Nicolini,
  \textit{Phys.\ Lett.\ B} {\bf 692}, 32 (2010).
  
\bibitem{BrM11} 
  E.~Brown and R.~B.~Mann,
  \textit{Phys.\ Lett.\ B} {\bf 694}, 440 (2011).


\bibitem{BMS08} 
  R.~Banerjee, B.~R.~Majhi and S.~Samanta,
 \textit{ Phys.\ Rev.\ D }{\bf 77}, 124035 (2008).


\bibitem{Sca99} 
  F.~Scardigli,
 \textit{ Phys.\ Lett.\ B} {\bf 452}, 39 (1999).
  
\bibitem{ChA03} 
  P.~Chen and R.~J.~Adler,
  \textit{Nucl.\ Phys.\ Proc.\ Suppl.\ } {\bf 124}, 103 (2003)

\bibitem{MaN11} 
  R.~B.~Mann and P.~Nicolini,
  \textit{Phys.\ Rev.\ D} {\bf 84}, 064014 (2011).

\bibitem{NiT11} 
  P.~Nicolini and G.~Torrieri,
  \textit{JHEP} {\bf 1108}, 097 (2011).

\bibitem{SmS13} 
  A.~Smailagic and E.~Spallucci,
  \textit{Int.\ J.\ Mod.\ Phys.\ D} {\bf 22}, 1350010 (2013).

\bibitem{SpS13} 
  E.~Spallucci and A.~Smailagic,
  \textit{J.\ Grav.\ } {\bf 2013}, 525696 (2013).

\bibitem{GaL09} 
  R.~Garattini and F.~S.~N.~Lobo,
  \textit{Phys.\ Lett.\ B} {\bf 671}, 146 (2009).


\bibitem{NiS10}
  P.~Nicolini and E.~Spallucci,
  \textit{Class.\ Quant.\ Grav.\ } {\bf 27} (2010) 015010.
  
  
\bibitem{NOS13} 
  P.~Nicolini, A.~Orlandi and E.~Spallucci,
 \textit{ Adv.\ High Energy Phys.\ } {\bf 2013}, 812084 (2013).



\bibitem{Riz06} 
  T.~G.~Rizzo,
 \textit{ JHEP} {\bf 09}, 021 (2006).


\bibitem{CaN08}
  R.~Casadio and P.~Nicolini,
 \textit{ JHEP }{\bf 11} (2008) 072.

\bibitem{Gin10} 
  D.~M.~Gingrich,
 \textit{ JHEP }{\bf 05}, 022 (2010)

\bibitem{NiW11} 
  P.~Nicolini and E.~Winstanley,
  \textit{JHEP} {\bf 11}, 075 (2011).

\bibitem{MNS12} 
  J.~Mureika, P.~Nicolini and E.~Spallucci,
  \textit{Phys.\ Rev.\ D }{\bf 85}, 106007 (2012).

\bibitem{BlN14} 
  M.~Bleicher and P.~Nicolini,
  ``Mini-review on mini-black holes from the mini-Big Bang,''
  arXiv:1403.0944 [hep-th].

\bibitem{MuN11} 
  J.~R.~Mureika and P.~Nicolini,
  \textit{Phys.\ Rev.\ D} {\bf 84}, 044020 (2011).
  

\bibitem{ANS07} 
  S.~Ansoldi, P.~Nicolini, A.~Smailagic and E.~Spallucci,
  \textit{Phys.\ Lett.\ B} {\bf 645}, 261 (2007)

\bibitem{SSN09} 
  E.~Spallucci, A.~Smailagic and P.~Nicolini,
 \textit{Phys.\ Lett.\ B} {\bf 670}, 449 (2009).

\bibitem{SmS10} 
  A.~Smailagic and E.~Spallucci,
  \textit{Phys.\ Lett.\ B} {\bf 688}, 82 (2010).


\bibitem{MoN10} 
  L.~Modesto and P.~Nicolini,
  \textit{Phys.\ Rev.\ D} {\bf 82}, 104035 (2010).
 
%
\bibitem{ArkaniHamed:1998rs} 
  N.~Arkani-Hamed, S.~Dimopoulos and G.~R.~Dvali,
  \textit{Phys.\ Lett.\ B} {\bf 429}, 263 (1998).
 %
\bibitem{ArkaniHamed:1998nn} 
  N.~Arkani-Hamed, S.~Dimopoulos and G.~R.~Dvali,
 \textit{ Phys.\ Rev.\ D} {\bf 59}, 086004 (1999).
%
\bibitem{Antoniadis:1998ig} 
  I.~Antoniadis, N.~Arkani-Hamed, S.~Dimopoulos and G.~R.~Dvali,
 \textit{ Phys.\ Lett.\ B} {\bf 436}, 257 (1998).
%
\bibitem{Randall:1999vf} 
  L.~Randall and R.~Sundrum,
 \textit{ Phys.\ Rev.\ Lett.} {\bf 83}, 4690 (1999).
  %
\bibitem{Randall:1999ee} 
  L.~Randall and R.~Sundrum,
 \textit{ Phys.\ Rev.\ Lett. } {\bf 83}, 3370 (1999).
%
\bibitem{Cavaglia:2002si} 
  M.~Cavaglia,
 \textit{ Int.\ J.\ Mod.\ Phys.\  A} {\bf 18}, 1843 (2003).
\bibitem{Kanti:2004nr} 
  P.~Kanti,
 \textit{ Int.\ J.\ Mod.\ Phys.\ A} {\bf 19}, 4899 (2004).
\bibitem{Cardoso:2012qm} 
  V.~Cardoso, L.~Gualtieri, C.~Herdeiro,
  U.~Sperhake, P.~M.~Chesler, L.~Lehner, S.~C.~Park
  and H.~S.~Reall {\it et al.},
\textit{  Class.\ Quant.\ Grav.} {\bf 29}, 244001 (2012).
\bibitem{Park:2012fe} 
  S.~C.~Park,
 \textit{ Prog.\ Part.\ Nucl.\ Phys.} {\bf 67}, 617 (2012).
 %
\bibitem{Calmet:2012mf} 
  X.~Calmet, L.~I.~Caramete and O.~Micu,
 \textit{ JHEP} {\bf 11}, 104 (2012).
  %
\bibitem{Arsene:2013nca} 
  N.~Arsene, X.~Calmet, L.~I.~Caramete and O.~Micu,
 \textit{ Astropart.\ Phys.} {\bf 54}, 132 (2014).
 %
\bibitem{Arsene:2013ria} 
  N.~Arsene, L.~I.~Caramete, P.~B.~Denton and O.~Micu,
  ``Quantum Black Holes Effects on the Shape of Extensive Air Showers,''
  arXiv:1310.2205 [hep-ph].
%
\bibitem{hawking}
S.W.~Hawking,
\textit{Nature} {\bf 248}, 30 (1974);
\textit{Comm. Math. Phys.} {\bf 43}, 199 (1975).
%
\bibitem{dimopoulos}
S.~Dimopoulos and G.~Landsberg,
\textit{Phys. Rev. Lett.} {\bf 87}, 161602 (2001).
 %
\bibitem{Banks:1999gd}
T.~Banks and W.~Fischler,
\textit{``A model for high energy scattering in quantum gravity,''}
arXiv:hep-th/9906038.
%
\bibitem{Giddings:2001bu}
S.B.~Giddings and S.D.~Thomas,
\textit{Phys.\ Rev.\  D} {\bf 65}, 056010 (2002).
%
\bibitem{charybdis}
C.~M.~Harris, P.~Richardson and B.R.~Webber,
\textit{JHEP} {\bf 08}, 033 (2003).
%
\bibitem{trunoir}
S.~Dimopoulos and G.~Landsberg,
{\em Black hole production at future colliders},
in {\it Proc.~of the APS/DPF/DPB Summer Study on the Future of Particle
Physics (Snowmass 2001)},
edited by N.~Graf, \textit{eConf} { C010630}, P321 (2001).
%
\bibitem{ahn}
E.J.~Ahn and M.~Cavaglia,
\textit{Phys. Rev. D} {\bf 73}, 042002 (2006).
%
\bibitem{catfish}
M.~Cavaglia, R.~Godang, L.~Cremaldi and D.~Summers,
\textit{Comput.\ Phys.\ Commun.} {\bf 177}, 506 (2007).
%
\bibitem{cha2}
G.~L.~Alberghi, R.~Casadio, A.~Tronconi,
\textit{J.\ Phys. G} {\bf 34 }, 767 (2007).
%
\bibitem{Alberghi:2006qr} 
  G.~L.~Alberghi, R.~Casadio, D.~Galli, D.~Gregori, A.~Tronconi and V.~Vagnoni,
  ``Probing quantum gravity effects in black holes at LHC,''
  hep-ph/0601243.
%
\bibitem{blackmax}
D.-C.~Dai, G.~Starkman, D.~Stojkovic, C.~Issever, E.~Rizvi, J.~Tseng,
\textit{Phys.\ Rev. D} {\bf 77 },   076007 (2008).
%
\bibitem{charybdis2}
J.A.~Frost, J.R.~Gaunt, M.O.P.~Sampaio, M.~Casals, S.R.~Dolan, M.A.~Parker, B.R.~Webber,
\textit{JHEP} {\bf 10 }, 014 (2009).
%
\bibitem{sampaio}
M.O.P.~Sampaio,
``Production and evaporation of higher dimensional black holes,''
\textit{Ph.D.~thesis}, http://www.dspace.cam.ac.uk/handle/1810/226741.
%
\bibitem{Landsberg:2006mm}
G.~Landsberg,
\textit{J. Phys. G} {\bf 32}, R337 (2006).
%
\bibitem{Harris:2004xt}
C.M.~Harris, M.J.~Palmer, M.A.~Parker, P.~Richardson,
A.~Sabetfakhri and B.~R.~Webber,
\textit{JHEP} {\bf 05}, 053 (2005).
%
\bibitem{Casanova:2005id}
A.~Casanova and E.~Spallucci,
\textit{Class. Quant. Grav.}  {\bf 23}, R45 (2006).
%
\bibitem{mfd}
R.~Casadio, B.~Harms and Y.~Leblanc,
\textit{Phys. Rev. D} {\bf 57}, 1309 (1998).
\bibitem{mfd1}
R.~Casadio and B.~Harms,
\textit{Phys. Rev. D} {\bf 58}, 044014 (1998).
\bibitem{mfd2}
R.~Casadio and B.~Harms,
\textit{Mod. Phys. Lett.} {\bf A14}, 1089 (1999).
%
\bibitem{entropy}
R.~Casadio, B.~Harms,
\textit{Entropy} {\bf 13 }, 502 (2011).
%
\bibitem{Koch:2005ks}
B.~Koch, M.~Bleicher and S.~Hossenfelder,
\textit{JHEP} {\bf 10}, 053 (2005).
%
\bibitem{Hossenfelder:2005bd}
S.~Hossenfelder,
\textit{Nucl. Phys. A} {\bf 774}, 865 (2006).
%
\bibitem{gingrich}
D.M.~Gingrich,
\textit{JHEP} {\bf 05}, 022 (2010).
%
\bibitem{Calmet:2008dg}
X.~Calmet, W.~Gong and S.D.H.~Hsu,
\textit{Phys.\ Lett.\  B} {\bf 668}, 20 (2008).
%
\bibitem{Calmet:2011ta} 
X.~Calmet, D.~Fragkakis and N.~Gausmann,
\textit{Eur.\ Phys.\ J.\ C} {\bf 71}, 1781 (2011).
%
\bibitem{Calmet:2012cn} 
X.~Calmet, D.~Fragkakis and N.~Gausmann,
``Non Thermal Small Black Holes,''
in {\em Black Holes: Evolution, Theory and Thermodynamics},
A.J.~Bauer and D.G.~Eiffel editors, Nova Publishers, New York, 2012
[arXiv:1201.4463].
%
%
\bibitem{Meade:2007sz}
P.~Meade and L.~Randall,
\textit{JHEP} {\bf 05}, 003 (2008).
%
\bibitem{Bellagamba:2012wz} 
L.~Bellagamba, R.~Casadio, R.~Di Sipio and V.~Viventi,
\textit{Eur.\ Phys.\ J.\ C} {\bf 72}, 1957 (2012).
%
\bibitem{Alberghi:2013hca} 
  G.~L.~Alberghi, L.~Bellagamba, X.~Calmet, R.~Casadio and O.~Micu,
  Eur.\ Phys.\ J.\ C {\bf 73}, 2448 (2013).
%
\bibitem{shiromizu}
T. Shiromizu, K. Maeda, M. Sasaki,
\textit{Phys. Rev. D} {\bf 62},  043523 (2000).
%
\bibitem{dadhich}
N.~Dadhich, R.~Maartens, P.~Papadopoulos, V.~Rezania,
\textit{Phys.\ Lett.} {\bf B487}, 1 (2000).
%
\bibitem{CH}
R.~Casadio and B.~Harms,
\textit{Int. J. Mod. Phys. A} {\bf 17}, 4635 (2002).
%
\bibitem{covalle}
R.~Casadio and J.~Ovalle,
\textit{Phys.\ Lett.\ B} {\bf 715}, 251 (2012).
\bibitem{covalle1}
R.~Casadio and J.~Ovalle,
\textit{``Brane-world stars from minimal geometric deformation, and black holes,''}
arXiv:1212.0409.
%
\bibitem{ida}
D.~Ida, K.~-y.~Oda and S.~C.~Park,
\textit{Phys. Rev. D} {\bf 67}, 064025  (2003)
[\textit{Erratum-ibid.\ D} {\bf 69} (2004) 049901].
%
\bibitem{ida1}
D.~Ida, K.~-y.~Oda and S.~C.~Park,
\textit{Phys. Rev. D} {\bf 71}, 124039 (2005).
\bibitem{ida2}
D.~Ida, K.~-y.~Oda and S.~C.~Park,
\textit{Phys. Rev. D} {\bf 73}, 124022 (2006).
%
\bibitem{kanti}
S.~Creek, O.~Efthimiou, P.~Kanti and K.~Tamvakis,
\textit{Phys.\ Rev.\ D} {\bf 75}, 084043 (2007).
\bibitem{kanti1}
S.~Creek, O.~Efthimiou, P.~Kanti and K.~Tamvakis,
\textit{Phys.\ Rev.\ D} {\bf 76}, 104013 (2007).
\bibitem{kanti2}
G.~Duffy, C.~Harris, P.~Kanti and E.~Winstanley,
\textit{JHEP} {\bf 09}, 049 (2005).
\bibitem{kanti3}
M.~Casals, P.~Kanti and E.~Winstanley,
\textit{JHEP} {\bf 02}, 051 (2006).
\bibitem{kanti4}
 M.~Casals, S.~R.~Dolan, P.~Kanti and E.~Winstanley,
\textit{JHEP} {\bf 03}, 019 (2007).
%
\bibitem{Casadio:2009sz}
  R.~Casadio, S.~Fabi, B.~Harms and O.~Micu,
  \textit{JHEP} {\bf 02} (2010) 079.
  %
\bibitem{Casadio:2010dq} 
  R.~Casadio, B.~Harms and O.~Micu,
 \textit{Phys.\ Rev.\ D} {\bf 82}, 044026 (2010).
%
\end{thebibliography}
\end{document}